\documentclass[fleqn,useAMS,usegraphicx, usenatbib]{mn2e}
\usepackage{aecompl}
\usepackage{mathptmx}
\usepackage[utf8]{inputenc}
\usepackage{amsmath}
\usepackage{amsfonts}
\usepackage{amssymb}
\usepackage{natbib}
\usepackage{graphicx}
\usepackage{aas_macros}
\usepackage[usenames,dvipsnames]{color}
\usepackage{xspace}
 \usepackage{fixltx2e} 
 \usepackage{sidecap} 

\usepackage[normalem]{ulem} 
\usepackage{enumitem} 

\DeclareMathOperator*{\dv}{d}
\newcommand{\hM}{_0}  
\newcommand{\eq}{_\mathrm{eq}}  
\newcommand{\gas}{_\mathrm{g}}  
\newcommand{\Prm}{P_\mathrm{m}} 
\newcommand{\Rey}{\mathrm{Re}}          
\newcommand{\Rem}{R_\mathrm{m}}          
\newcommand{\Remcr}{R_\mathrm{m,cr}}       
\newcommand{\Breg}{\overline{B}}

\newcommand{\ee}{\mathrm{e}}  
\newcommand{\B}{B}  				
\newcommand{\BB}{\overline{B}}  	
\newcommand{\BBv}{\overline{\vect{B}}}  	
\newcommand{\bb}{b}  	
\newcommand\vect[1]{\pmb{#1}}			
\newcommand{\crit}{_\mathrm{c}}  
\newcommand{\MF}{_\mathrm{MF}}  
\newcommand{\cm}{\,{\rm cm}}    
\newcommand{\km}{\,{\rm km}}    
\newcommand{\pc}{\,{\rm pc}}    
\newcommand{\kpc}{\,{\rm kpc}}  

\newcommand{\cgsdensity}{\,{\rm g}\,{\rm cm}^{-3} }      
\newcommand{\msun}{\text{M}_\odot}

\newcommand{\s}{\,{\rm s}}      
\newcommand{\yr}{\,{\rm yr}}    
\newcommand{\Myr}{\,{\rm Myr}}  
\newcommand{\Gyr}{\,{\rm Gyr}}  

\newcommand{\kms}{\km\s^{-1}}    

\newcommand{\muG}{\,\mu{\rm G}} 
\newcommand{\mkG}{\,\mu{\rm G}} 

\newcommand{\erg}{\,{\rm erg}}  
\newcommand{\K}{\,{\rm K}}      

\newcommand{\glf}{\textsc{galform}\xspace}

\definecolor{webgreen}{rgb}{0,.5,0}
\definecolor{webbrown}{rgb}{.6,0,0}
\definecolor{purple}{rgb}{0.5,0,.5}

\usepackage[pdfborder={0 0 0.1}]{hyperref}
\hypersetup{
pdfauthor={L.~F.~S.~Rodrigues (orcid.org/0000-0002-3860-0525), A.~Shukurov (orcid.org/0000-0001-6200-4304), A.~Fletcher, C.~M.~Baugh (orcid.org/0000-0002-9935-9755)},
pdftitle={Galactic magnetic fields and hierarchical galaxy formation}
}

\begin{document}
\title[Magnetic fields and galaxy formation]
{Galactic magnetic fields and hierarchical galaxy formation}
\author[L.~F.~S.~Rodrigues, A.~Shukurov, A.~Fletcher,
C.~M.~Baugh]{L.~F.~S.~Rodrigues$^{1}$\thanks{Email:
luiz.rodrigues@newcastle.ac.uk}, A.~Shukurov$^{1}$,
A.~Fletcher$^{1}$ and C.~M.~Baugh$^{2}$
\\
$^{1}$School of Mathematics and Statistics, University of Newcastle, Newcastle
upon Tyne, NE1 7RU, UK.\\
$^{2}$Institute for Computational Cosmology, Department of Physics, University
of Durham, South Road, Durham, DH1 3LE, UK.
}

\date{Accepted 2015 April 09. Received 2015 April 09; in original form 2015
January 04}
\pagerange{\pageref{firstpage}--\pageref{lastpage}}
\pubyear{2015}

\maketitle
\begin{abstract}
A framework is introduced for coupling the evolution
of galactic magnetic fields sustained by the mean-field dynamo
with the formation and evolution of galaxies in cold dark matter cosmology.
Estimates of the steady-state strength of the large-scale and turbulent magnetic
fields from mean-field and fluctuation dynamo models are used together with galaxy
properties predicted by semi-analytic models of galaxy formation for a population of
spiral galaxies.
We find that the field strength is mostly controlled by
the evolving gas content of the galaxies.
Thus, because of the differences in the implementation of the star formation law,
feedback from supernovae and ram-pressure stripping, each of the galaxy
formation
models considered predicts a distribution of field strengths with unique
features.
The most prominent of them is the difference in typical magnetic
fields strengths obtained for the satellite and central galaxy populations
as well as the typical strength of the large-scale magnetic field in
galaxies of different mass.
\end{abstract}

\label{firstpage}
\section{Introduction}
Magnetic fields have been detected in all spiral galaxies when observed at
sufficient sensitivity and resolution \citep{BBMSS96,BW13}. The most informative
observational tracers of magnetic
fields are partially polarised synchrotron emission and Faraday rotation.
Polarised radio emission, together with significant Faraday rotation,
indicate the presence of a component of the galactic magnetic field
ordered on scales comparable to the resolution of the observation,
of order a few hundred parsecs for nearby galaxies, and exceeding the integral
turbulent scale which is of order $50$--$100\pc$.
The random (turbulent) magnetic field
$\vect{\bb}$ is usually stronger than the large-scale component $\BBv$, so
the degree of polarisation does not typically exceed 10--20 per cent.
An average total field strength in nearby galaxies is
$\B=(\BB^2+\bb^2)^{1/2}\approx 9\muG$ with $\bb/\BB\simeq1$--3 \citep{BBMSS96},
ranging
from about $4\muG$ in M31 \citep{FBBS04} to  $15\muG$ in M51 \citep{FBSBH11}.

Magnetic fields contribute significantly to the structure and evolution
of the interstellar medium (ISM) and the host galaxy.
They affect the accretion of gas by 
{dark matter haloes \citep{Rodrigues2010},
as well as the outflows and inflows in galaxies that have already formed \citep{Bir09}.
Magnetic fields also influence outflows since they affect the multi-phase
structure of the interstellar gas as they confine hot gas bubbles produced by 
supernovae \citep{FMZ91,KBST99,SSNGB04,HT06}.
The magnetic contribution to the overall structure of galactic gas discs is
at least as important as that from other sources of interstellar pressure, i.e. 
thermal, turbulent and from cosmic rays, as all of them
are of comparable magnitude \citep{P79book,B87,BC90,FS01}.
Half of the total interstellar pressure is thus due to non-thermal contributions, 
and, hence, magnetic fields directly affect the mean gas density. 
In turn, this affects significantly the star 
formation rate; it is perhaps surprising that the role of magnetic 
fields and cosmic rays in galaxy evolution has avoided attention 
for so long \citep{BBT15}. Magnetic fields also regulate star formation locally by
controlling the collapse and fragmentation of molecular clouds \citep{ML09,PBKM11,C12}.
Magnetic fields contribute to interstellar gas dynamics not only
directly but also by confining cosmic rays \citep{BBDP90}. The latter are
effectively weightless and so are capable of driving galactic outflows (winds
and fountains)
\citep{
       BK00,EZBMRG08,UPSNES12,BAKG13} thus providing negative feedback 
       on star formation in galactic discs \citep{VCB05,PPJ12}.
}

Future radio telescopes such as the Square Kilometre
Array\footnote{
http://www.skatelescope.org}
will allow a dramatic increase in the sensitivity and angular resolution of
radio
observations to permit not only detailed studies of nearby galaxies but also
reliable and extensive measurements of galactic magnetic fields at higher
redshifts
\citep{Gaensler2004}.
As we demonstrate here, the evolution of galactic magnetic fields is sensitive
to a variety of poorly constrained parameters in galactic evolution models and
can thus provide a sensitive diagnostic of the physical processes involved in
galaxy formation.

The most successful theory for the production of magnetic
fields observed in galaxies is the turbulent dynamo \citep[for reviews,
see][]{BBMSS96,S07}. Two types of dynamo are expected to be
important: the fluctuation dynamo, which produces a small-scale, random magnetic
field of coherence lengths comparable to the turbulent scale in the plasma,
and the mean-field dynamo, which produces a large-scale magnetic field
ordered on scales greater than the turbulent scale. The details of the two
dynamos are briefly discussed in
Section~\ref{sec:B}.

There have been several attempts to understand the behaviour of large-scale
magnetic fields in an evolving galaxy but they have been limited by the quality
of galactic formation and evolution models available. \citet{BPSS94} suggested that
the seed magnetic field for the large-scale dynamo can be produced in
a protogalaxy or young galaxy by the fluctuation dynamo acting on a short time
scale \citep[see also][]{KCOR97,Arshakian2009,SBSAKBS10,SFSBK12,SBS14},
and thus the observed galactic magnetic fields can be generated
in 1--2\,Gyr for typical galactic properties.
The seed magnetic field for the fluctuation dynamo can be produced by
many mechanisms \citep{DN13} such as battery effects in the first
generation of stars \citep{HTKOIO05} or in a rotating protogalaxy \citep{MR72}.

The only evolutionary effect included by \citet{BPSS94}
is a variation of the thickness of the galactic ionised gas layer with time.
\citet{Arshakian2009,Arshakian2011} discuss dynamo action at various
stages of galaxy formation at a qualitative level with reference to more recent
models of galaxy formation and evolution.

Here we quantitatively model the large-scale galactic magnetic fields produced
by a
mean-field dynamo in the framework of specific hierarchical galaxy formation
models,
with the aim of predicting the statistical properties of galactic magnetic
fields in a
large sample of galaxies at different redshifts. We employ semi-analytic
models (SAMs) of galaxy formation based on simple physical prescriptions
\citep[for a
review, see][]{BaughReview}.
These models produce synthetic catalogues of galaxies that are able to reproduce
a
wide range of observables
(e.g., the galaxy luminosity and stellar mass functions). Starting from the
stellar
and gaseous masses, disc sizes,
star formation rates and circular velocities, obtained from SAMs for late-type
galaxies,
we calculate magnetic fields using dynamo theory. The galaxy formation models
remain uncertain regarding the details of the physical processes involved
and their relative importance. In particular, the nature and intensity of the
feedback
of star formation on the galactic discs, which is in turn sensitive to the form
and strength of the
interstellar magnetic field, remains unclear.

We derive the dependence of the strength of magnetic field
on the galactic mass, its evolution with redshift, and the statistical
distribution of magnetic field strengths
in galactic samples. Remarkably, different SAMs lead to distinct predictions
regarding galactic magnetic fields. Thus, magnetic fields observed in
high-redshift galaxies via their polarisation and Faraday rotation can help to
refine galaxy formation models.

{An alternative approach to galaxy formation and evolution in a cosmological
context involves the numerical solution of the magnetohydrodynamic (MHD)
equations for
interstellar gas in the (evolving or static) gravity field obtained from
$N$-body
simulations. Gas dynamics is often simulated with adaptive mesh refinement 
(or a particle based technique such as Smooth Particle Hydrodynamics) in 
order to achieve higher spatial resolution in denser regions. In the best MHD
simulations
available now, the highest resolution is 20--300\,pc \citep{WA09,PMS14}. The
driving
of interstellar turbulence by supernovae is neglected or parameterised.
Moreover,
the turbulent dynamo action responsible for both large-scale and turbulent
interstellar
magnetic fields obviously needs a fully resolved turbulent flow to be modelled
accurately. As a result, the processes responsible for the large-scale magnetic
field
cannot be captured at all, and the simpler fluctuation dynamo is controlled by
random
flows at unrealistic scales. Even so, such results should be 
treated with care as turbulent magnetic fields are known to be sensitive to 
flows well within the inertial range of interstellar turbulence at scales of a
fraction of
parsec \citep{BS05}. Thus, magnetohydrodynamic simulations of galaxy formation
and
evolution need to be complemented with semi-analytic models coupled with a
mean-field
description of the large-scale magnetic field and models of the turbulent
fields. The
advantages of this approach include the opportunity to explore the parameter
space and
clarify the role of various physical effects. This opportunity is particularly
attractive given that many important physical processes are already heavily 
parameterised in the hydrodynamic and magnetohydrodynamic models.
}

The paper is organised as follows.
In Section~\ref{sec:SAM}, we discuss semi-analytic models of galaxy formation
and evolution and Section~\ref{sec:B} describes the generation of galactic
magnetic fields by dynamo mechanisms.
Our results for the distribution and evolution of galactic magnetic fields
can be found in Section~\ref{sec:results}, together with a discussion on the
systematic connection between the gas content of galaxies and magnetic fields,
the role of magnetic helicity diffusion, the redshift evolution of galactic
magnetic fields emerging from our models 
{
and the dependence of magnetic field 
on the star formation rate.
}
Our conclusions are presented in Section~\ref{sec:conclusions}.
The appendices provide further details of the models.

\section{Galactic evolution}
\label{sec:SAM}

To obtain the galaxy properties needed to compute the strength of galactic
magnetic fields we use semi-analytic models of galaxy
formation (SAMs). These break down galaxy formation and evolution into a
set of differential equations,
each of which models a separate physical process, including:
the merger history of dark matter haloes;
the radiative cooling of gas inside these haloes and the subsequent
formation of galactic discs;
star formation in galactic discs;
galaxy mergers driven by dynamical friction and bursts of star formation
associated with them;
the feedback due to supernovae and AGNs
and the chemical enrichment of stars and gas. (For reviews on these topics,
see \citealt{BaughReview,BensonReview}.)

We use three versions of the Durham SAM, \glf \citep{Cole2000, Bower2006}, the
version presented by \citet{Baugh2005}, hereafter referred to as BAU, that
of \citet{Lagos2012}, referred to as LAG and that of \citet{Font2008},
hereafter FON. All models reproduce the galactic luminosity
function at redshift $z=0$ in the $K$ and $b_J$ bands.
The BAU version of \glf successfully reproduces the counts and redshift
distribution of galaxies selected according to their sub-millimetre
luminosity, and the abundances of Lyman-break galaxies \citep{Lacey2011} and
Lyman~$\alpha$ emitters \citep{Orsi2008}.
The FON version provides a better match to the observed colours of satellite
galaxies.
The LAG version reproduces the atomic and molecular hydrogen mass
functions \citep{Lagos2011gas}, the $K$-band luminosity function at redshifts
$z=1$ and $2$ \citep{Lagos2011sfl} and the CO(1-0) luminosity function.
Both LAG and FON are derived from the model of \citet{Bower2006} and predict
the properties of the AGN population \citep{Fanidakis2011,Fanidakis2012}.

The choice of these models allows us to assess the variability of the
predictions
between state-of-art SAMs and the impact of these differences on the the
galactic
magnetic fields inferred from them.

\subsection{Semi-analytic models of galaxy formation and evolution}

The SAMs explored differ in various aspects, briefly described here. Further
details can be found in the original papers referred to in the previous
section.

\subsubsection{Cosmological model and dark matter merger trees}

The models differ in their assumed cosmologies and dark matter merger trees.
The BAU model adopts a present-day cosmological constant energy
density parameter $\Omega_\lambda=0.7$, a matter density parameter
$\Omega_m=0.3$, a Hubble parameter $h=0.7$ and a spectrum normalisation
$\sigma_8=0.93$. The LAG and FON models adopt the cosmological parameters of
the Millennium Simulation \citep{Millennium}, $\Omega_\lambda=0.75$,
$\Omega_m=0.25$, $h=0.73$ and $\sigma_8=93$.

The dark matter (DM) merger trees used in the LAG and FON models are extracted
from the Millennium $N$-body cosmological simulation \citep{Millennium}, which
has a halo mass resolution of $1.72\times10^{10}h^{-1}\msun$.
In the BAU model, the merger trees are built using the Monte-Carlo
procedure described by \citet{Cole2000}, with the minimum final halo mass for
the merger trees at $z=0$ set to $5\times10^9h^{-1}\msun$.

\subsubsection{Star formation and initial mass function}
The models differ in their treatment of star formation in galaxy discs
(quiescent star formation). In the BAU model, the galactic star formation rate
is calculated as the ratio of the total mass of cold gas to the characteristic
star formation time scale, $\text{SFR}=M_\mathrm{g}/\tau_\star$, with
\begin{equation}
\label{eq:BAU_SFL}
\tau_\star =
\tau_{\star0} \left(\frac{V\hM}{200\kms}\right)^{\alpha_\star}\,,
\end{equation}
where $V\hM$ is the circular velocity at the half-mass radius of the disc,
$\tau_{\star0}=8\Gyr$ and $\alpha_\star=-3$. This prescription reproduces
the observed gas fraction--luminosity relation at $z=0$.

In the FON model, the star formation time scale uses the parameterisation
\begin{equation}
\label{eq:FON_SFL}
\tau_\star =
\frac{\tau_{\rm disc}}{\epsilon_\star}
\left(\frac{V\hM}{200\kms}\right)^{\alpha_\star}\,,
\end{equation}
where $\tau_{\rm disc}=r\hM/V\hM$ is the disc dynamical time-scale and the free
parameters $\epsilon_\star=0.0029$ and $\alpha_\star=-1.5$ were chosen mainly
to reproduce the galaxy luminosity function at $z=0$ \citep{Bower2006}.

In the LAG model, the star formation rate per unit area is assumed to be
proportional
to the surface density of molecular hydrogen in the galactic disc.
The amount of H$_2$ is obtained from the empirical relation of
\citet{BlitzRosolowsky} which relates the fraction of molecular hydrogen
to the pressure in the mid-plane of the galactic disc.

The star formation prescription used in LAG has no free parameters in
the sense that they are calibrated against observations of star formation
rates and gas surface densities in nearby galaxies.
The parameters are in effect measured and are only allowed to span a narrow
range of
values consistent with observational errors, narrower than the
ranges within which other model parameters are typically allowed to vary.

The stellar initial mass function (IMF) is assumed to be
universal in LAG and FON, of the form proposed by \citet{KennicutIMF}.
BAU also assumes the Kennicutt IMF for quiescent star formation, but
a top-heavy IMF is used for the starbursts.

\subsubsection{Starbursts}

In all models, galaxy mergers can trigger bursts of star formation,
which transfer gas (and possibly stars) from the disc to the spheroidal
component, where this gas participates in star formation.
{The merger events are instantaneous in the models  and lead to
either an elliptical galaxy (for a merger of galaxies of similar masses) or to
a corresponding increase in the galaxy mass if the merging satellite galaxy has
a mass much smaller than the central galaxy.}

In LAG and FON, material is transferred to the spheroidal component, triggering
a starburst {also in the case} when the galactic disc becomes dynamically
unstable according to the bar stability criterion of \citet*{ELN}. The BAU
model does not allow for any disc instabilities.

\subsubsection{Stellar and AGN feedback}
\label{sec:feedback}
The models adopt similar parameterisations of the  outflow due to
supernovae and stellar winds, with a mass loss rate from the disc that
depends on the star formation rate (a measure of the energy input into the ISM)
and some power of the circular velocity of the disc (a measure of the depth of
the potential well). The mass expelled is reincorporated into the disc
at a later time.
In the BAU model, this
happens once the dark matter halo has doubled its mass, whereas in LAG and in
FON, this occurs after a time $t = t_\text{dyn}/\alpha_{\rm rh}$, where
$t_\text{dyn}=r_\text{virial}/V_\text{virial}$ is the dynamical time of the
halo and $\alpha_{\rm rh}= 1.26$.

On the other hand, the models differ significantly in their assumptions about
the processes that regulate the bright end of the galactic luminosity function.
In BAU, galactic super-winds
are assumed to eject material from haloes
that have low circular velocity (with no subsequent reincorporation), at a rate
proportional to the star formation rate. This reduces the baryon fraction in
more massive haloes, which themselves are not directly subject to superwinds,
thereby reducing the rate at which gas cools in these haloes.
In LAG and FON, the growth of supermassive black holes in the centre
of the galaxies is modelled and associated with an AGN feedback
contribution.
The cooling flow is interrupted once the black hole's Eddington
luminosity exceeds a multiple of the cooling luminosity.

\subsubsection{Ram-pressure stripping}
\label{sec:ram-pressure}
When a galaxy becomes a satellite galaxy (after the merger of dark matter
haloes), it looses part or all of its hot gas halo to the central system by
ram-pressure stripping.

In BAU and LAG, this is modelled by transferring all of the hot gas halo
of the satellite to the central galaxy, i.e., one assumes that the hot gas is
completely and instantaneously stripped.
In the FON, ram-pressure stripping is modelled more accurately, without
assuming the instantaneous stripping of the gas. Instead, the
prescription obtained from hydrodynamical simulations
of \citet{McCarthy2008} is followed.

\subsection{Late-type galaxies and their properties}
\begin{table*}
\caption{Summary of the quantities used and notation.}
\centering
\begin{tabular}{llll}
\hline
Type of parameter
&Notation    &Meaning    &Defined in or\\

                    &
&                    &the value adopted\\
\hline
Galactic properties obtained   & $M\gas$ & Cold gas mass of a galaxy \\
from the semi-analytic models  & $M_\star$ & Stellar mass of a galaxy \\
of galaxy formation            & $r_{50,\text{out}}$ & Half-mass radius
of the galactic disc\\
                & $V\hM$ & Circular velocity of the galactic disc at the
    half-mass radius\\
                & SFR & Star formation rate in the disc \\
\hline
Quantities estimated in this paper
             & $\overline{h}$ & Average scale height of the galactic disc
&Eq.~\eqref{eq:h})\\
             & $\overline{\rho}$ & Average gas density in the galactic disc
&Eq.~\eqref{eq:rho} \\
             & $r\hM$ & Corrected half-mass disc radius
    &Eq.~\eqref{eq:correctR}\\
             & $\Omega$ & Angular velocity of the disc
&Eq.~\eqref{eq:Omega}\\
             & $S$ & Maximum rotational shear
 &Eq.~\eqref{eq:S}\\
             & $v_\text{ad}$ & Local outflow speed
    &Eq.~\eqref{eq:vad}\\
\hline
Adopted parameters 
             & $l_0$ & Characteristic length scale of the turbulence
&$0.1\kpc$\\
                         & $v_0$ & Root-mean-square gas velocity dispersion in
the disc    &$10\kms$ \\
            & $\alpha$ & Number of contributions to the interstellar pressure
& 4 (Eq.~\ref{eq:P})\\
                         & $R_\kappa$ &  Ratio of turbulent diffusivities of the
mean helicity and large-scale magnetic field    &0.3 (Eq.~\ref{eq:Rkappa}) \\

\hline
Computed quantities
             & $R_u$ & Outflow magnetic Reynolds number
&Eq.~\eqref{eq:Ru}\\
             & $D$ & Dynamo number     &Eq.~\eqref{eq:D})\\
             & $D\crit$ & Critical dynamo number     &Eq.~\eqref{eq:Dc}\\
             & $\Breg$ & Steady-state large-scale magnetic fields
                         strength                         &Eq.~\eqref{eq:B} \\
             & $b$ &  Steady-state random magnetic field strength
&Eq.~\eqref{eq:b}\\
\hline
\label{tab:summary}
\end{tabular}
\end{table*}

Magnetic fields ordered at the galactic scale only occur in late-type
galaxies (since they have significant rotation), so we focus on these
galaxies in what follows.
We classify a galaxy as late-type if its bulge stellar mass accounts for less
than half the total galaxy mass.

\begin{figure}
 \centering
 \includegraphics[width=\columnwidth]{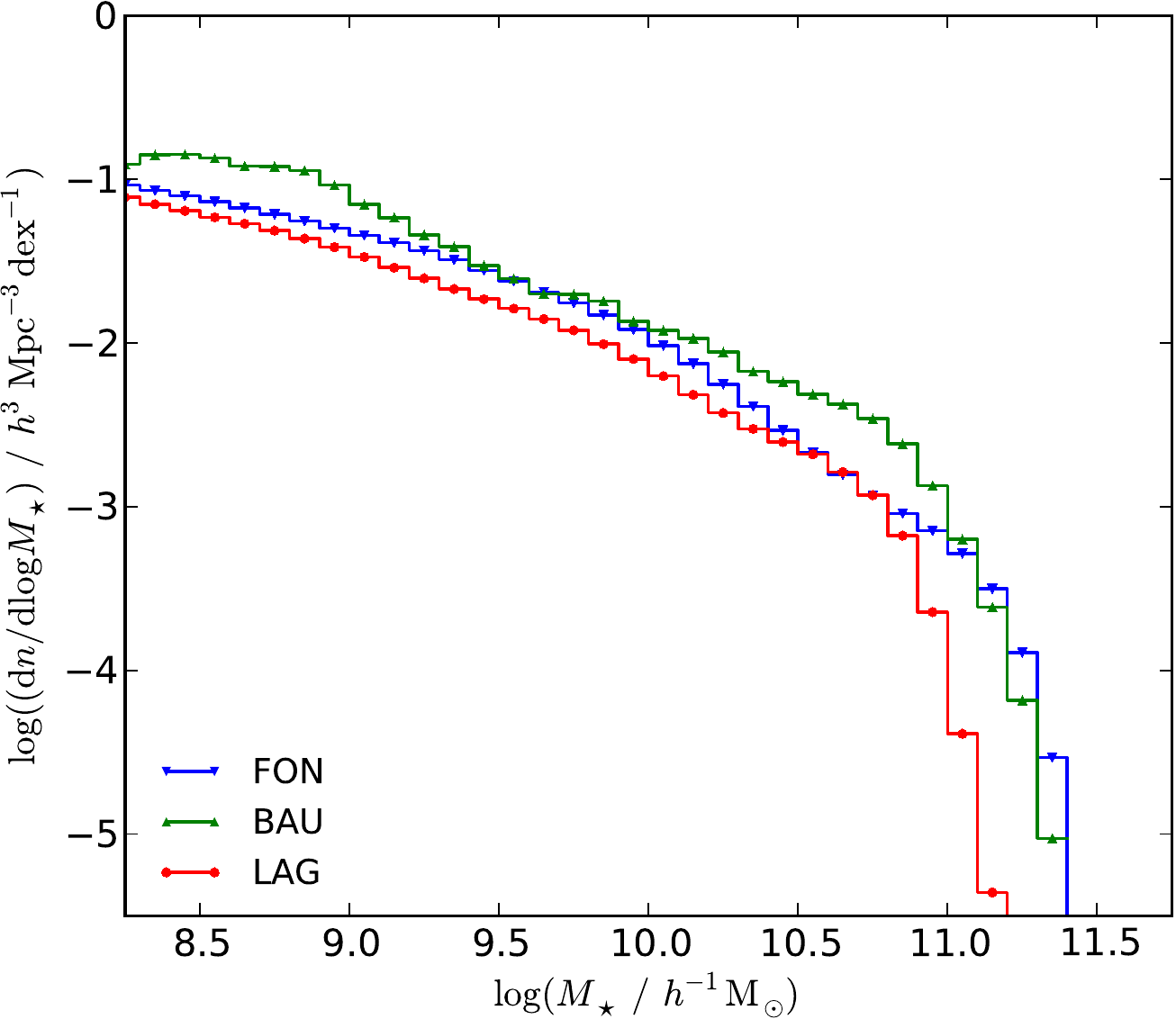}
 \caption{The late-type galaxy stellar mass function of the SAMs used.
          Different colours correspond to different SAMs as labelled.
          }
 \label{fig:smf}
\end{figure}

Details of the galaxy formation models affect the distribution of galaxy
stellar
masses. In Fig. \ref{fig:smf}, this is shown for the explored models.

\subsubsection{Interstellar turbulence}\label{sec:IT}
Interstellar gas is vigorously driven into a turbulent state by the
injection of kinetic energy,
mostly from supernova explosions (a general consequence of star formation), but
also from stellar winds and other, less important sources
\citep{M_LK04,ES04a,ES04b}. The root-mean-square
(rms) turbulent speed is close to the sound speed in the warm gas
($T\simeq10^4-10^5\K$) at $v_0\simeq10\kms$ and exhibits weak
variations within and between spiral galaxies \citep{Tamburro2009}.
The turbulent scale
$l_0\simeq0.1\kpc$ is controlled by the size of a supernova remnant when
its expansion velocity decreases to the sound speed in the ambient
warm gas: the expanding supernova shell drives flows in its wider environment
only from this stage onwards.

The small variability in the turbulent speed may seem surprising
as one might expect a dependence of $v_0$ on the star formation rate.
Indeed, detections of such a dependence through H\,\textsc{i} observations
have been reported \citep{Dib2006}. However, both the interpretation of such
observations and the connection between star formation rate and turbulence
are far from straightforward.
Apart from driving turbulent motions, supernova activity is responsible
for the hot phase of the interstellar gas, and the filling factor of the hot
phase increases with SFR \citep{AvillezBreitschwerdt2004}.
The energy released by supernovae is distributed among several such channels
(including radiative cooling, acceleration of cosmic rays, etc.), and it is not
obvious that an
enhanced star formation rate would necessarily lead to a larger turbulent
velocity. The fact that the turbulence observed in the warm interstellar gas
is transonic is likely not to be a coincidence, but rather a result of
non-linear feedback
between star formation and turbulence. If the turbulent driving becomes stronger
because of an increase in the SFR, the turbulence initially becomes supersonic,
leading to a rapid dissipation of kinetic energy into heat in the shocks, so
that the turbulent kinetic energy is reduced, being converted into thermal
energy of the warm gas (the speed of sound in the hot gas is of order
$100\kms$, so the warm gas absorbs most of the extra kinetic energy). As a
result, the gas temperature increases until the turbulence becomes transonic and
the system reaches a new (quasi-)steady state. However, the gas cools
radiatively, on a cooling time scale (of a few $10^3\,$yr for temperatures near
$10^4\K$, assuming solar metallicity)
that is shorter than the typical time scale of variations in the star formation
rate (of order $10^6\yr$). Thus, the warm gas can adjust itself to the varying
star formation rate remaining at a nearly constant equilibrium temperature of
$T=10^4$--$10^5\K$ which is known to be rather insensitive to the heating rate
because of the efficient
cooling. As a result, the extra energy supplied by the supernovae is more
plausibly lost as radiation and, more importantly here, increases the
fractional volume of the hot gas. This enhances the associated outflow
from the gas disc but does not increase the turbulent velocity.
In other words, the turbulent velocity in the warm gas is controlled by its
sound speed, i.e., by the balance of heating and cooling which, ultimately,
is determined by its chemical composition (the metal abundance).

Outflows (fountains or winds) driven by supernova activity entrain significant
amounts of warm and cold gas \cite[e.g., ][]{GSFSM134a,MOMQ14}, so it is not
surprising that observations at a relatively low resolution (1\,kpc or larger)
can pick up a broad range of velocities dominated by outflows rather than the
turbulent motions in the warm gas.

To conclude, we keep the turbulent velocity and scale fixed and independent
of the star formation rate at $v_0=10\kms$ and $l_0=0.1\kpc$. However, the
outflow speed is a sensitive function of the star formation rate in our model
(Section~\ref{GO}).

\subsubsection{Derived galactic quantities}

\begin{figure*}
 \centering
  \includegraphics[width=\textwidth]{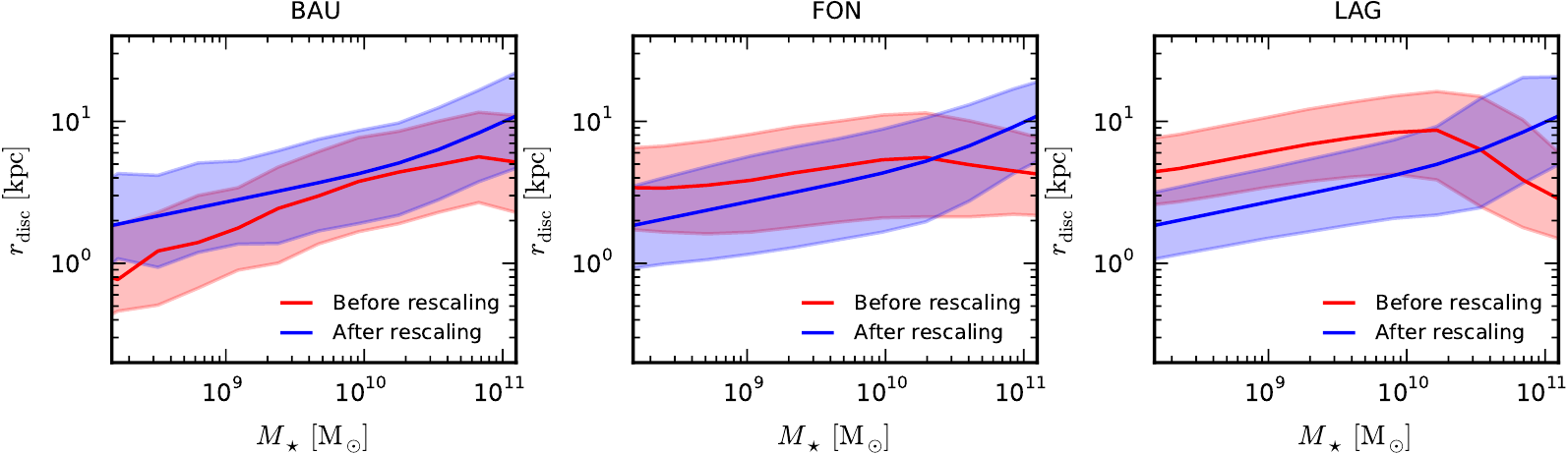}
\caption{
Disc size--stellar mass relation for the three SAMs used here. The \emph{red}
curves and corresponding shaded areas show to medians and the range between the
15th and 85th percentiles for the raw model output. The \emph{blue} curves and
corresponding shaded areas show the same data after the rescaling to make the
medians match the fit to observational data given by equation \eqref{eq:MSR}.
Each panel shows a different model as labelled.
}
 \label{fig:rdisc}
\end{figure*}

The calculation of the magnetic field in a galaxy that is described in
Section \ref{sec:B} requires knowledge of the average gas density,
$\overline{\rho}$, and the average disc height, $\overline{h}$. In this section
we derive simple expressions for these two quantities as functions of galaxy
properties output by SAMs (see Table \ref{tab:summary} for a summary of the
notation
and quantities involved).

Assuming that the surface density of the disc of the galaxy is well described by
an exponential profile,  the height
of a gaseous disc in hydrostatic equilibrium is given by (see
Appendix~\ref{ap:h}
for a derivation)
\begin{equation}
 h(r)=\frac{\alpha v_0^2 r_s^2}{3\,G(M_\star + M_\text{gas})}
 \,\ee^{r/r_s}\,,\label{eq:hr}
\end{equation}
where $v_0$ is the turbulent velocity of the disc gas, $r_s$ is the radial scale
length of the disc and $\alpha$ is the number of pressure contributions to the
support of the gaseous disc.
We use $\alpha=4$, assuming equal contributions from thermal, turbulent,
cosmic-ray and magnetic pressures.

Equation \eqref{eq:hr} leads to a divergent galaxy volume
unless it is truncated. We choose the half-mass radius as the galactic boundary,
$r\hM = \xi r_s$ with $\xi \approx 1.678$, so that the volume of the galaxy
follows as
\begin{equation}
\label{eq:vol}
 V = 2\pi \int_0^{r\hM} h(r) r \,\dv r = 2\pi\lambda r\hM^2h(0)\,,
\end{equation}
where
\[
 \lambda = \int_0^1 x \ee^{\xi x} \,\dv x\approx 1.645 \,.
\]

The average disc scale height can be defined as
\begin{equation}
 \overline{h} = \frac{V}{2\pi r\hM^2}=
 \frac{\alpha v_0^2 r\hM^2}{3 G\left(M_\star+M_\text{gas} \right)} \xi
 \label{eq:h}\,,
\end{equation}
which yields
\begin{align}
 \overline{h} \simeq & 287\pc
\left(\frac{r\hM}{10\kpc}\right)^2
\left(\frac{v_0}{10\kms}\right)^2\\
                      &\times \frac{\alpha}{4}
 \left(\frac{M_\star + M_\text{gas}}{10^{11}\msun}\right)^{-1}
 .\nonumber
\end{align}

The average gas density is then obtained as
\begin{equation}
\overline{\rho}=\frac{M_\text{gas}}{2\,V} =
\frac{G M_\text{gas} (M_\text{gas} +M_\star)}{\alpha v_0^2 r\hM^4}
\frac{3\xi^2}{4\pi\lambda}\label{eq:rho}\,,
\end{equation}
or
\begin{align}
\overline{\rho}\simeq& 1.06 \times 10^{-23}\cgsdensity
\left(\frac{\alpha}{4}\right)^{-1}
\left(\frac{r\hM}{10\kpc}\right)^{-4}
\\ &\times
\left(\frac{M_\text{gas}}{10^{10}\msun}\right)
\left(\frac{M_\text{gas}+M_\star}{10^{11}\msun}\right)
\left(\frac{v_0}{10\kms}\right)^{-2}.
\nonumber
\end{align}

{The factor $1/2$ in equation \eqref{eq:vol} allows for the fact that
$r_0$ in equation \eqref{eq:vol} is the half-mass radius, thus enclosing half
of
the total mass.
}
\subsubsection{Disc size}
Reproducing the observed sizes of galactic discs is a long
standing problem in semi-analytic
models of galaxy formation \citep[e.g.][]{Gonzalez2009}.
Because of the strong dependence of the derived quantities
on the disc size, we have renormalised the disc half-mass radii so that the
medians of the predicted half-mass radii match the fit obtained by 
\citet{Dutton2011} for the
observed relation between the half-mass radius and the stellar mass of SDSS
galaxies at $z=0$,
\begin{equation}
r_{50,\text{emp}}(z=0) = r_D \left(\frac{M_\star}{M_D}\right)
\left[
\frac{1}{2}+\frac{1}{2}\left(\frac{M_\star}{M_D}\right)^{c_3}
\right]^{(c_2-c_1)/c_3}\,,
 \label{eq:MSR}
\end{equation}
where
\[
(c_1,c_2,\log_{10}M_D,\log_{10}r_D,c_3)
=
(0.2, 0.46, 10.39, 0.75,1.95)\,.
\]

Thus, the corrected half mass radius of a disc galaxy in a particular mass bin
is
related to the original disc size in the output of the SAM, $r_{50,\text{out}}$
through
\begin{equation}
\label{eq:correctR}
 r\hM(z) = r_{50,\text{out}}(z)
\frac{r_{50,\text{emp}}(z=0)}{r_{50,\text{median}}(z=0)}
\end{equation}

This procedure corrects the median values of the disc radii in each mass
bin, preserving the dispersion of disc sizes predicted by the models and
the redshift evolution of the mass-size relation. In Fig. \ref{fig:rdisc} we
show the original and corrected size-mass relations
for each SAM.

\subsubsection{Rotation and velocity shear}

An estimate of the disc angular velocity, $\Omega$, is necessary in the
modelling of galactic magnetic fields. This is computed at the
half-mass radius, $r\hM$, using
\begin{equation}
\Omega = \Omega\hM=\frac{V\hM}{r\hM}\label{eq:Omega}\,,
\end{equation}
where $V\hM$ is the disc's circular velocity at $r=r\hM$.

The next quantity to compute is the shear, $S=r\,\partial\Omega/\partial r$.
We employ the maximum value of $S$ in our calculations.
It can be shown that the maximum shear in a purely exponential disc
is related to the angular velocity through
\begin{equation}
  S \approx - 0.76\, \Omega\hM\,.\label{eq:S}
\end{equation}

\subsubsection{Galactic outflows}\label{GO}

The mean-field dynamo, responsible for the generation of a large-scale galactic
magnetic field, relies on the removal of small-scale fields (and their
magnetic helicity) away from the dynamo region in order to
saturate in a steady state where the field has a strength comparable to that
observed, of several microgauss \citep{BS05}.
As discussed by \citet{Shukurov2006}, in galaxies this can be achieved by
the advection of magnetic fields by the outflow of the hot gas (and its
entrained colder components) from the disc{, as well as by turbulent diffusion
\citep[][and references therein]{KMRS00}}.

While the SAMs have internal prescriptions to compute the amount of gas
which outflows from the disc due to SNe powered galactic winds, the gas carried
by these winds is thought to be removed completely from the galaxy, and becomes
available for reincorporation into the hot halo only after time scales larger
than dynamical time of the dark matter halo.
These strong winds, however, correspond  only to a fraction of the total
outflow.

The outflow relevant for the magnetic field 
evolution is
the one associated with galactic fountains, which leads to the removal
of hot magnetised gas from the mid-plane of the disc but with a possibly very
short re-accretion time-scale. This is not considered explicitly in the SAMs. We
model this process as follows (see also \citealt{Lagos2013}).

Supernovae tend to cluster in OB associations, regions of active star formation.
The large, energetic bubbles of hot gas (superbubbles) produced by dozens of
supernovae and stellar winds in an OB association can more readily break out
from the galactic disc.
In the framework of the superbubble model of \citet{MacLowMcCray1988},
the speed of the shock front at the top of an expanding superbubble at it
breaks-out is given by (see Appendix~\ref{ap:OB} for derivation)
\begin{equation}
v_\text{sh} =  4\kms
\left(\frac{n_0}{1\cm^{-3}}\right)^{-1/3}
\left(\frac{h}{1\kpc}\right)^{-2/3},
\label{eq:vadsingle}
\end{equation}
where $n_0$ is the mean number density of the gas in the disc. Note that 
we have assumed that the OB associations can be treated as identical (i.e.,
they
share a common equivalent mechanical luminosity and the SN rate within them is 
approximately the same).

The interesting quantity, however, is the average of $v_\text{sh}$ over the
whole galaxy, $\overline v_\text{sh}$, taking into account the \emph{rate of
occurrence} of OB associations. This can be estimated as follows
\begin{equation}
 \overline v_\text{sh} = v_\text{sh}
\frac{A_\mathrm{OB}}{A_\text{gal}}\label{eq:vadratio}
 \text{ ,}
\end{equation}
where  
{$A_\text{gal}=\pi\,r_0^2$ is the surface}
area of the galaxy. The total area filled with OB associations is
\begin{equation}
 A_\mathrm{OB} = N_\mathrm{OB} \pi (2 h)^2\,,
\end{equation}
{
since the radius of an OB association at the break out is about $2h$ 
\citep{MacLowMcCray1988}.
}
The number of OB associations can be obtained from the frequency of SNe
occurrence in OB associations, $\nu_\mathrm{SN,OB}$, and the rate of supernovae
occurrence in a single OB association,$\nu_\mathrm{SN,1OB}$,
\begin{equation}
 N_\mathrm{OB} = \frac{\nu_\mathrm{SN,OB}}{\nu_\mathrm{SN,1OB}}\,.
\end{equation}

The frequency of SN occurrence in OB associations is approximately $70$ per
cent of the total SN frequency \citet{TenorioTaglerBodenheimer1988}, i.e.
$\nu_\mathrm{SN,OB} = f_\mathrm{OB} \nu_\mathrm{SN}$
with $f_\mathrm{OB}\approx0.7$. The overall SNe frequency relates to the star
formation
rate through $\nu_\mathrm{SN} = \eta_\mathrm{SN}\times$SFR, where 
$\eta_\mathrm{SN} = 9.4\times 10^{-3}\msun^{-1}$ for Kennicutt's IMF 
\citep{KennicutIMF}.

The frequency of supernovae within a single OB association can
be 
{found from}
the number of SNe that typically occur in one OB 
{association,}
$N_\mathrm{SN, 1OB}\approx 40$ \citep{Heiles1987}, and the typical
{lifetime} of an OB association is $t_{OB} =3\times 10^6\,$yr, i.e.,
$\nu_{SN,1OB} {\simeq} t_\mathrm{OB}^{-1}\,{N_\mathrm{SN, 1OB}}$.
Thus, we find that the number of OB associations
in a galaxy at a given epoch is related to its star formation rate through
\begin{align}\label{eq:NOB}
 N_\mathrm{OB} \approx&\,\,  490
 \left(\frac{\text{SFR}}{\msun\,\text{yr}^{-1}}\right)
 \left(\frac{N_\mathrm{SN,1OB}}{40}\right)^{-1}    \nonumber\\
 &\times 
 \left(\frac{t_\mathrm{OB}}{3\times10^{-3}\Gyr}\right)
 \left(\frac{f_\mathrm{OB}}{0.7}\right)\,.
\end{align}

Using equations \eqref{eq:vadsingle}--\eqref{eq:NOB} one obtains
\begin{align}
\label{eq:vout}
\overline v_\text{sh} =&\,\,  {1.5  \kms}
	\left(\frac{\text{SFR}}{\msun\,\text{yr}^{-1}}\right) 
		\left(\frac{n_0}{1 \,\text{cm}^{-3}}\right)^{-1/3}\nonumber\\
 &\times 
 \left(\frac{\overline{h}}{200
\pc}\right)^{\!4/3}\left(\frac{r\hM}{5\kpc}\right)^{-2}\,.
\end{align}
{This is the gas speed in (and immediately behind) the shock front
of a supperbubble. At the time scales involved, the magnetic field can be
assumed
to be frozen into the gas and thus is lost from the disc together with the gas.
Therefore, the quantity of interest is the mass-averaged speed, $v_\text{ad}$, 
such that the surface density of mass loss from the disc is $\overline{\rho}
v_\text{ad}$:}
\begin{equation}
 v_\text{ad} = \overline{v}_\text{sh}\frac{\rho_h}{\overline\rho}\,,
\label{eq:vad}
\end{equation}
where 
{$\rho_h \approx  1.7 \times 10^{-27} \cgsdensity$ is the density of the
hot gas} 
and $\rho$ is the average interstellar gas density.

\section{Models of galactic magnetic fields}
\label{sec:B}
A partially ordered magnetic field $\mathbf{B}$ in the turbulent interstellar
gas
can be conveniently represented as the sum of a  large-scale,
$\overline{\mathbf{B}}$, and a fluctuating, $\mathbf{b}$, components,
\[
\mathbf{B}=\overline{\mathbf{B}}+\mathbf{b}\,, \qquad
\overline{\mathbf{b}}=\mathbf{0}\,,
\]
where a bar denotes the ensemble or any other suitable average
\citep[see][for a discussion of averaging procedures]{GSSFM13}.

The growth time of the large-scale magnetic field due to the mean-field
turbulent dynamo \citep{Mo78,RSS88} can be estimated as \citep{JCBS14}
\begin{equation}\label{tauMF}
{
\tau\MF\simeq2\frac{h^2}{\eta}(D\crit-D)^{-1/2}\,,
\quad
|D|>|D\crit|\,,
}
\end{equation}
where the magnetic diffusivity $\eta$, dominated by the turbulent contribution,
is estimated from mixing-length theory as
\[
\eta\simeq \tfrac{1}{3}l_0 v_0\,,
\]
with $l_0$ and $v_0$ the turbulent scale and speed, respectively, 
{the dynamo}
number $D$ (that quantifies the strength of the dynamo action) can
be written using equations~\eqref{eq:D} and \eqref{eq:alpha0} as
\[
D\simeq \left(\frac{hl_0}{\eta} \right)^2\Omega S\,,
\]
with $S$ being the rotational shear rate,
{and $D\crit$ is a critical value of the dynamo number such that
magnetic field can be maintained only if $D< D\crit$ ($D<0$ so long 
as $\Omega$ decreases with $r$).}
We note that $h^2/\eta$ is the turbulent diffusion time across a layer
of a scale height $h$.
{Equation~\eqref{tauMF} is quite accurate for the range of $|D|$ of
interest in applications to spiral galaxies \citep[see the discussion
in][]{JCBS14}.
For a flat rotation curve, $S=-\Omega$, 
and $|D|\gg|D\crit|$, which is true in the inner parts of most galaxies,}
we obtain
{$D\simeq -10\left(h\Omega/v_0\right)^2$ and}
\begin{align}\label{tauMFD}
\tau\MF&\simeq\frac{2h}{l_0\Omega}\\
	&\simeq4\times10^8\yr
\left(\frac{h}{0.5\kpc} \right)
\left(\frac{\Omega}{25\,\text{km/s/kpc}}\right)^{-1}
\left(\frac{l_0}{0.1\kpc} \right)^{-1},
\end{align}
normalised to the parameter values corresponding to the Solar vicinity of the
Milky Way.
In the inner parts of slowly rotating galaxies and in outer regions of normal
disc galaxies, $\tau\MF$ is shorter than the galactic evolution
time scale $\tau_\mathrm{e}\simeq10^9\yr$.

Random magnetic fields grow even faster due to fluctuation dynamo action
(note that this dynamo action is distinct from the more commonly discussed
mean-field dynamo), on a time scale shorter than the
turnover time of the largest turbulent eddy,
$\tau_0\simeq l_0/v_0\simeq10^7\yr$.

The exponential growth on the time scales specified above halts (the dynamo
action is said to saturate) as soon as the Lorentz force becomes comparable to
other forces on the relevant length scale, of order 1\,kpc, for the large-scale
magnetic field and
a fraction of the turbulent scale for the random field \citep{BS13}. After that,
magnetic fields remains in
a statistically steady state. The magnitudes of $\overline{\mathbf{B}}$ and
$\mathbf{b}$ in this state are
discussed further in this section. A recent review and a suite of formulae
describing nonlinear mean-field dynamos can be found in \citet{Chamandy2014}.

Thus, as a first approximation, we can assume that: (i)~galactic magnetic fields
adjust themselves instantaneously to the evolving galactic environment
(i.e, $\tau\MF,\tau_0\ll\tau_\mathrm{e}$); and (ii)~they always remain in a
statistically steady state of a saturated dynamo. These assumptions are
obviously crude.
However, they are sufficient for a first exploration of the effects of magnetic
fields on galaxy evolution if the exploration then continues to include the
effects of finite dynamo time scales. The latter will be our subject elsewhere.

\subsection{Random magnetic fields in spiral galaxies}
\label{sec:Bsmall}
Any random flow of electrically conducting fluid is a dynamo (i.e., it amplifies
a seed magnetic field exponentially fast) provided the magnetic Reynolds number
{due to Ohmic magnetic diffusivity,}
$\Rem=l_0v_0/\eta$, exceeds a certain critical value, $\Remcr$
of order 100 in an incompressible flow. This type of a dynamo is known as the
\textit{fluctuation dynamo} \citep{ZRS90,SCMM02,BS05}.

The growth time of the random magnetic field in a \textit{vortical\/} random
velocity field of a scale $l$ is as short as the kinematic time at that scale,
$\tau\simeq l/v(l)$ with $v(l)$ the rms random speed at the scale $l$.
In a turbulent flow with a sufficiently shallow power spectrum, $\tau$
is shorter on smaller scales. For example, $\tau\propto l^{2/3}$
in a flow with the Kolmogorov
spectrum $v(l)\propto l^{1/3}$. Therefore, the magnetic energy spectrum peaks
at small scales during the exponential growth (kinematic) stage, and then
plausibly
settles to a form similar to that of the kinetic energy spectrum in the
statistically steady state (a saturated, nonlinear dynamo). The eddy turnover
time in spiral galaxies,
where $l_0\simeq0.1\kpc$ and $v_0\simeq10\kms$,
is as short as $\tau_0\simeq l_0/v_0\simeq10^7\yr$ even at the energy
scale of the interstellar turbulence.
Thus, the fluctuation dynamo can rapidly produce random magnetic fields in the
ISM \citep{S07}. 

{Compressibility hinders the fluctuation dynamo, reducing the growth rate
of the random magnetic field. In the extreme case of sound-wave turbulence,
the longitudinal nature of the fluid motions reduces the probability of
three-dimensional
twisting and folding of magnetic lines, an essential element of the
amplification mechanism 
(as in Zeldovich's stretch--twist--fold dynamos -- \citealt*{ZRS83,ZRS90}).
As a result, the growth time of the rms magnetic field is of order 
$\tau\simeq\mathcal{M}^{-4}l_0/v_0$ for $\mathcal{M}\ll1$, where $\mathcal{M}$ 
is the Mach number \citep{KRS85}. 
A generic compressible turbulence inherits this feature, producing a higher
threshold
value of $\Rem$ and  slower growth of magnetic energy
\citep{HBM04,FCSBKS11,GSFSM134a}.
We note, however, that the velocity field in such flows is a mixture
of solenoidal and  compressible parts \citep{MS96}, with their relative
contributions depending
on the Reynolds number, Mach number, numerical resolution, etc.\
\citep{HBM04,MB06}.
\citet{HBM04} find, in their simulations of dynamo action in a compressible
random flow, that $\Remcr$ increases from about 35 at $\mathcal{M}=0$ to 60 at
$\mathcal{M}=1$ and 80 at $\mathcal{M}=2.5$ (for the magnetic Prandtl number 
$\Prm=\Rem/\Rey=1$ with $\Rey$ the Reynolds number) but suggest that $\Remcr$ 
may vary little as the Mach number 
increases further, especially for a large magnetic Prandtl number (as in the 
interstellar gas). \citet{FCSBKS11} find $\tau\propto\mathcal{M}^{-1/3}$ 
for $10<\mathcal{M}<20$ in their simulations of isothermal compressible random
flows.}

{However, slower growth and a higher dynamo threshold in a compressible random
flow,
as compared with incompressible flows, may not in fact represent a practical
problem
since the value of the magnetic Reynolds number is usually very high in
astrophysical
plasmas, and the turbulent kinematic time scale $l_0/v_0$ is
very short in comparison with any global time scale in galaxies. What is more
important is the steady-state magnitude of the magnetic energy density relative
to the kinetic energy density of turbulence. According to \citet{FCSBKS11},
isothermal compressible random flows produce lower magnetic energy density
$E_\mathrm{m}=b^2/8\pi$
relative to the kinetic energy density $E_\mathrm{k}=\tfrac12\rho v_0^2$,
with $E_\mathrm{m}/E_\mathrm{k}\simeq10^{-2}\text{--}10^{-3}$ at
$2<\mathcal{M}<20$
than a purely solenoidal velocity field. Thus, gas compressibility  is 
detrimental to the fluctuation dynamo in both the kinematic and saturated
regimes.}

The magnetic field produced by the fluctuation dynamo is intermittent, and can
be described as a statistical ensemble of magnetic flux ropes and sheets
whose coherence size (length or radius of curvature) is of the order of the
flow correlation length, $l_0$, whereas the rope thickness is, in the
kinematic dynamo, of the order of the resistive scale, $l_0\Rem^{-1/2}$,
\citep{ZRS90}. \citet{WBS07} show that flux ropes become progressively
dominant as $\Rem$ increases in the kinematic regime.

Kinematic fluctuation dynamos are well understood, but the nonlinear,
statistically-steady state remains, to some extent, controversial.
Simulations of fluctuation dynamos in driven random flows suggest that
magnetic energy density within the ropes is close to equipartition
with the kinetic energy density \citep{BS05},
\begin{equation}
B\eq=v_0(4\pi\rho)^{1/2}\,.\label{eq:Beq}
\end{equation}
A widely accepted model for the saturation of the fluctuation dynamo by
\citet{S99}
\citep[see also ][]{SSH06}, suggests that, in the steady state, the magnetic
Reynolds number is renormalised to its critical value, so that the
thickness of magnetic ropes is estimated as
$d\simeq l_0\Remcr^{-1/2}\simeq0.1 l_0$ for $\Remcr=100$ independently of
$\Rem$ \citep[see, however, ][]{SCHMM02}. For modest effective magnetic
Reynolds numbers, $\Rem\simeq\Remcr$, magnetic sheets may be predominant.
Then the volume filling factor $f_B$ of the magnetic structures can be estimated
assuming that there is one magnetic sheet per correlation cell of the
turbulent flow:
\[
f_B=\frac{d}{l_0}\simeq\Remcr^{-1/2}\simeq0.1\text{ ,}
\]
so that the root-mean-square magnetic field follows as
\begin{equation}
b\simeq f_B B\eq = f_B\,v_0(4\pi\rho)^{1/2}\text{ .}\label{eq:b}
\end{equation}
However, magnetic field outside the dominant magnetic structures can still
contribute
significantly to the observables, such as 
{the random Faraday rotation measure}
\citep{BS13},
so that this estimate should be applied judiciously.

\subsection{Regular magnetic fields in spiral galaxies}
\label{sec:Blarge}

The generation of large-scale magnetic fields in galaxies is described by
galactic dynamo
theory using the concept of a mean-field dynamo \citep[see][for a
review]{BBMSS96,S07}.
Dynamo action in a rotating, stratified galactic gas layer is produced by the
joint action of the helical
turbulent motion 
(via the so-called $\alpha$-effect) and the differential
rotation of the galactic disc. These two effects are quantified, respectively,
by two dimensionless parameters,
\[
 R_{\alpha} =\frac{\alpha_0 h}{\eta}
 \quad
 \text{and}
 \quad
 R_{\omega} = \frac{S h^2}{\eta}\,.
\]
The widely used $\alpha\omega$-dynamo approximations applies where
$|R_\omega|\gg R_\alpha$.
In this approximation, dynamo action is controlled solely by their product,
known as the dynamo number,
\begin{equation}
\label{eq:D}
D=R_\alpha R_\omega\,.
\end{equation}
The magnitude of the $\alpha$-effect can be obtained from the order-of-magnitude
estimate
\begin{equation}\label{eq:alpha0}
\alpha_0 \simeq \frac{l_0^2 \Omega}{h}\,.
\end{equation}

It is not quite clear how the mean-field dynamo enters a nonlinear,
steady state \citep[for a review, see][]{BS05}. In the most detailed and
physically motivated theory available, the growth of the large-scale magnetic
field is limited by the conservation of magnetic helicity, so its steady-state
strength depends on the rate at which the mean magnetic helicity of the
random magnetic field is removed from the localisation region of the
large-scale magnetic field \citep{DSGB13}.
In galaxies, this can be accomplished by galactic winds and fountain flows
\citep{Shukurov2006,Sur2007}. In this case an additional dimensionless
parameter enters the picture, quantifying the \emph{advection} of
magnetic helicity out of the galactic disc
\begin{equation}
 R_u = \frac{v_\text{ad} h}{\eta}\,.\label{eq:Ru}
\end{equation}

Another mechanism that contributes to the partial removal of the small-scale
magnetic helicity is its turbulent diffusion \citep{KMRS00,KMRS02,KMRS03}.
Allowing for this effect introduces another dimensionless number, the
diffusivity ratio of the mean current helicity, $\kappa$, and the mean magnetic
field, $\eta$:
\begin{equation}
R_\kappa = \frac{\kappa}{\eta}\,.\label{eq:Rkappa}
\end{equation}
\citet{Mitra2010} obtained $R_\kappa\approx 0.3$ independent of the Reynolds
number.
\medskip

\begin{figure*}
 \centering
 \includegraphics[width=\textwidth]{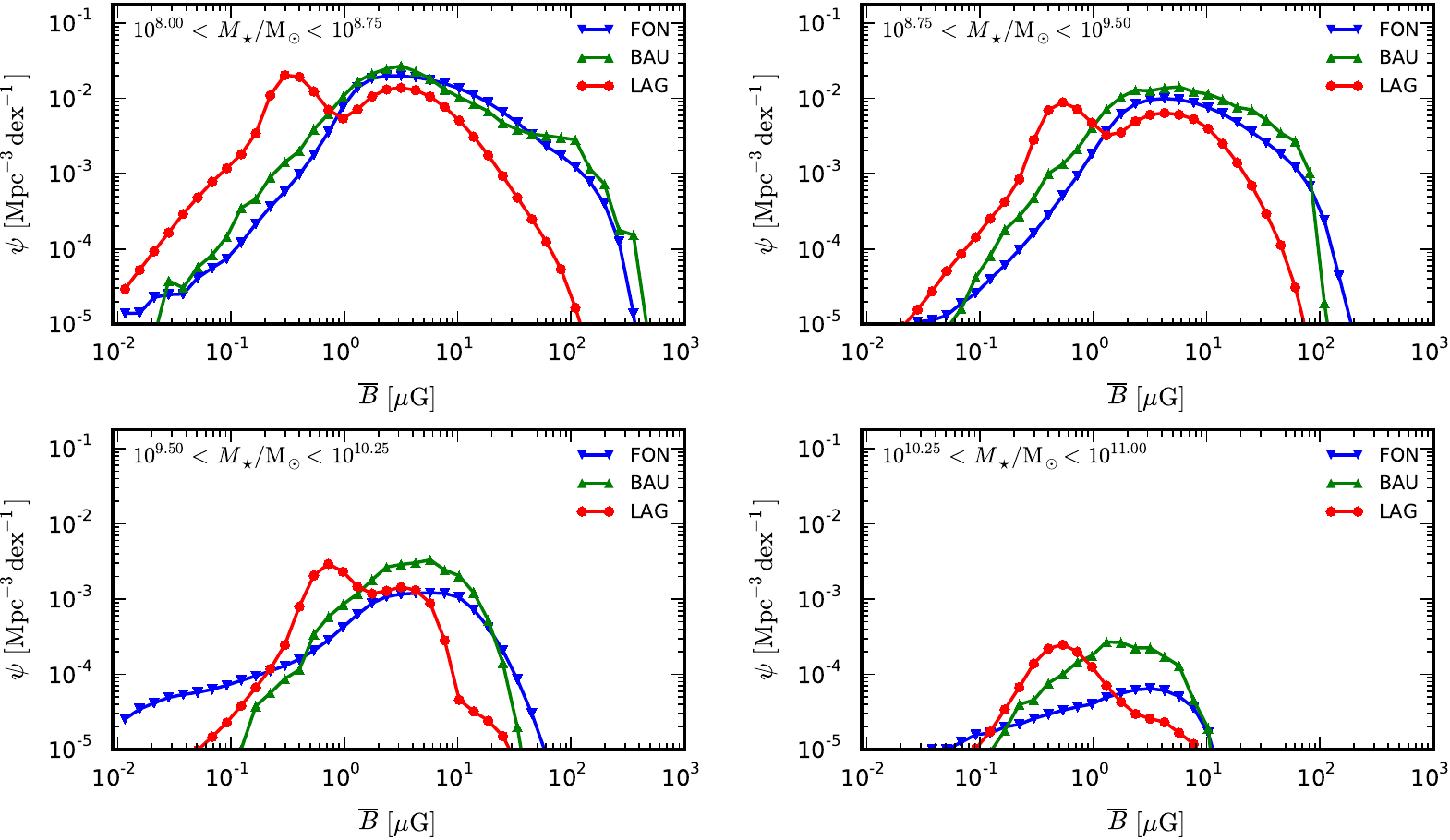}
 \caption{
The predicted magnetic strength function (MSF), equation \eqref{eq:msf} of the 
large-scale magnetic fields of spiral galaxies at $z=0$  in different
SAMs, as indicated by the key.
Each panel corresponds to  a different galactic stellar mass interval, indicated
in the top left of each frame.
 The points correspond to binned model results.
}
 \label{fig:msf}
\end{figure*}

To summarise this model \citep[see][for details]{Chamandy2014}, a disc galaxy
has an active mean-field dynamo (i.e., it is capable of amplifying and
sustaining a large-scale magnetic field) if the the magnitude of the dynamo
number, equation~\eqref{eq:D}, exceeds that of its critical value,
\begin{equation}
D\crit = -\left(\frac{\pi}{2}\right)^5 \left( 1+\frac{R_u}{\pi^2}\right)^2\,.
\label{eq:Dc}
\end{equation}
For $|D|<|D\crit|$, the large-scale magnetic field decays on a time scale
of order $h^2/\eta$.
Otherwise, $\BBv$ is exponentially amplified up to the steady-state strength
\begin{equation}
 \BB^2\simeq
 \tfrac12 \zeta(p) B\eq^2 \left(\frac{l_0}{h}\right)^2
 \left(\frac{D}{D\crit} -1\right)\left(R_u+\pi^2 R_\kappa\right)\,,
 \label{eq:B}
\end{equation}
where the magnetic pitch angle, defined as
$p\equiv\arctan\left(\BB_r/\BB_\phi\right)$
in terms of the cylindrical magnetic field components, can be expressed in term
of the previous quantities as
\begin{equation}
p=\arctan\left(
\frac{1}{R_\omega} \left|\frac{2\,D\crit}{\pi^{1/2}}\right|
\right),
\end{equation}
with $B\eq$ defined in equation~\eqref{eq:Beq}, and
\begin{equation}
\zeta(p) = \left(1-\frac{3\sqrt{2}}{8}\cos^2 p\right)^{-1} \,.
\end{equation}

\section{Results and discussion}
\label{sec:results}

In this section we identify and interpret robust features of galactic magnetic
fields obtained from the galaxy formation models. Our results refer to
statistical
properties of large-scale and random magnetic fields in large samples of
galaxies. We found that magnetic properties are different in galaxies of
different masses and our results are presented for
characteristic galactic mass ranges, where we have calculated the mass using
only the stellar content.

The statistical properties of galactic magnetic fields are convenient to
describe in terms of the number density of galaxies with a particular
magnetic field strength per logarithmic interval of the filed strength; in the
case of the large-scale magnetic field, this variable, referred to as the
magnetic strength function (MSF), is defined as
\begin{equation}
\label{eq:msf}
  \psi\left(\Breg\right) = \frac{\dv n}{\dv\,\log\Breg}\,.
\end{equation}

\subsection{The large-scale magnetic field}
Figure~\ref{fig:msf} shows the distribution of magnetic field strength, the MSF,
for the large-scale field $\Breg$ in the local Universe ($z=0$), each panel
displaying a different galactic mass interval.
The form of the MSF is sensitive to the galaxy formation model
and, in each model, to the mass interval.

The typical magnetic field strengths are similar in the FON and BAU models,
$0\lesssim\log_{10}(\Breg/ 1\muG)\lesssim1$.
Except for the highest mass interval, the MSF of the LAG model is clearly
bimodal, with the first peak in the interval $0.2-0.8\muG$ and the second
close to the what is predicted by the other models.

Lower mass galaxies can have the strongest large-scale magnetic fields. Strong
large-scale magnetic fields are increasingly suppressed with increasing mass and
for galaxy masses $M_\star>10^{10.25}\,\msun$ there is a negligible number
density of galaxies with large-scale magnetic fields stronger than $\sim
10\muG$. {The reason for the suppression, discussed in more
detail in section \ref{sec:gascontent}, is the increase in
the outflow speed in these galaxies, associated with their higher star
formation rates.}

Another factor that affects Fig. \ref{fig:msf} there is the overall decrease in 
the number density of galaxies with the galactic  mass. 
This is expected from the shape of the galaxy
stellar mass function  of late-type galaxies, shown in
Fig.~\ref{fig:smf}. However, as it will be discussed later, the decrease is
also intensified by the {relatively} small fraction of {massive} 
galaxies with active dynamos.

\begin{figure}
 \centering
 \includegraphics[width=\columnwidth]{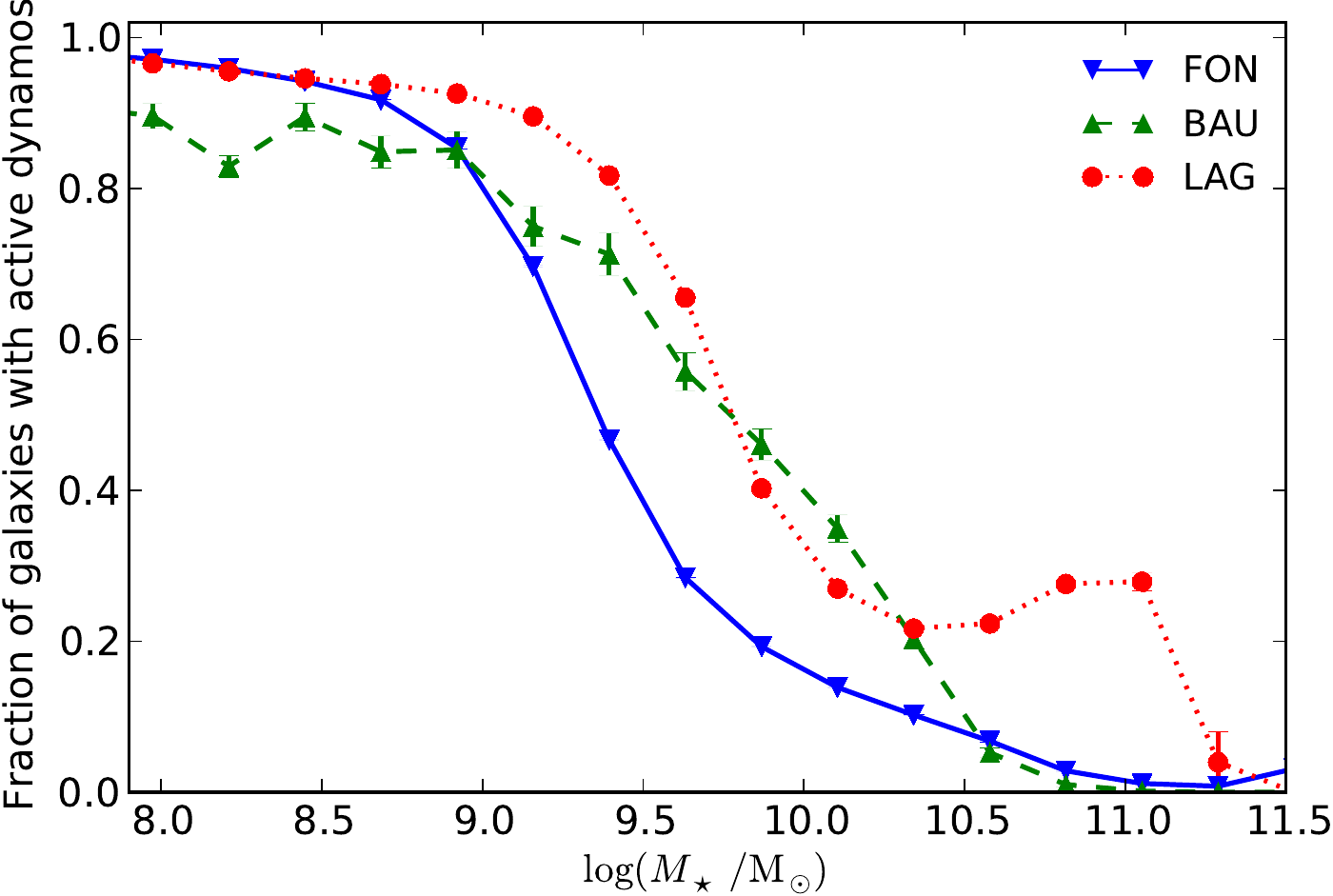}
 \caption{
          The fraction of galaxies with active dynamos (i.e. $|D|>|D_c|$) at
          $z=0$ as a function of galaxy mass in each model, as shown in the
          legend.
         }
 \label{fig:fractionsA}
\end{figure}
\begin{figure}
 \centering
\includegraphics[width=\columnwidth]{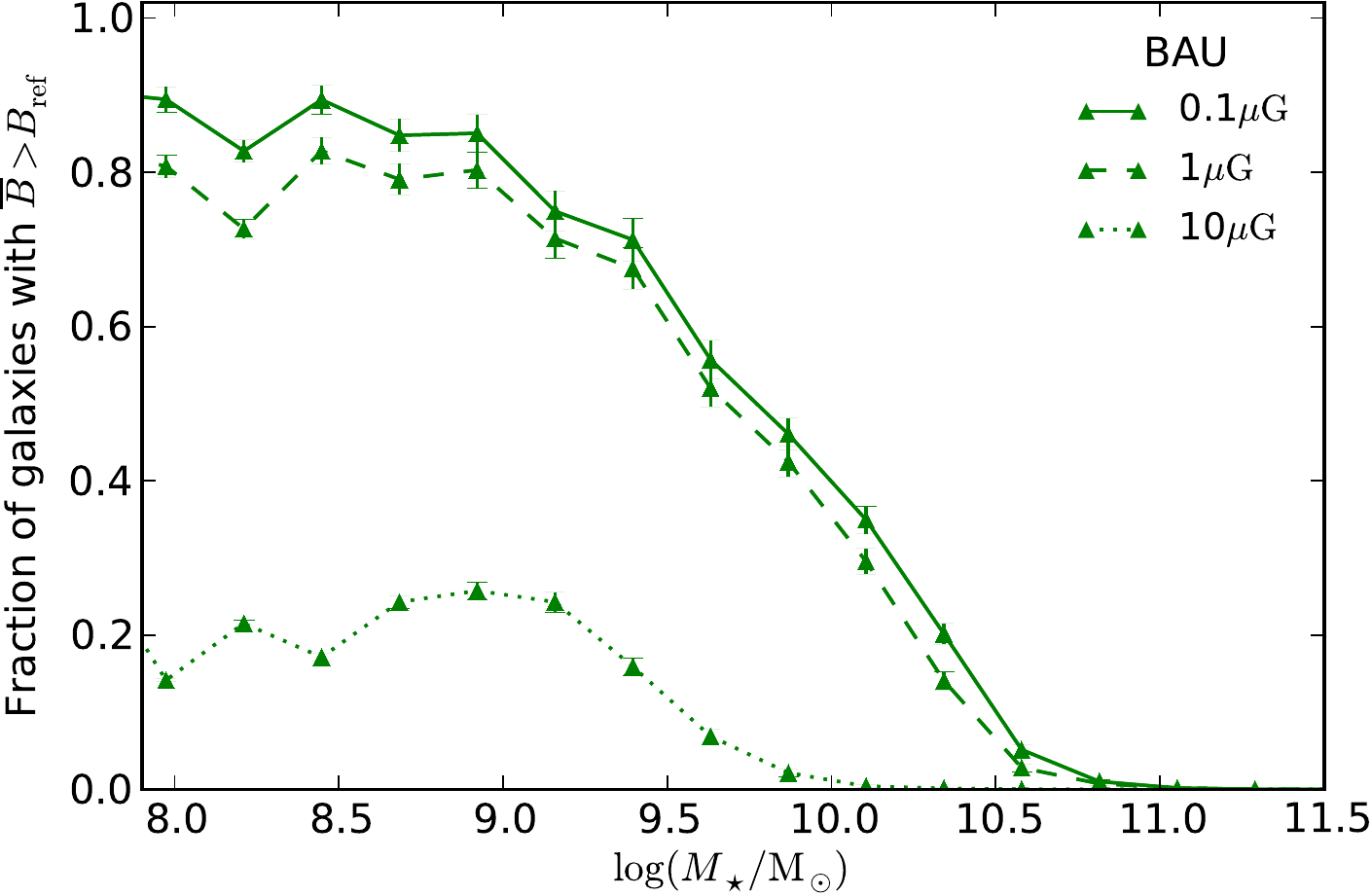}
\\
\includegraphics[width=\columnwidth]{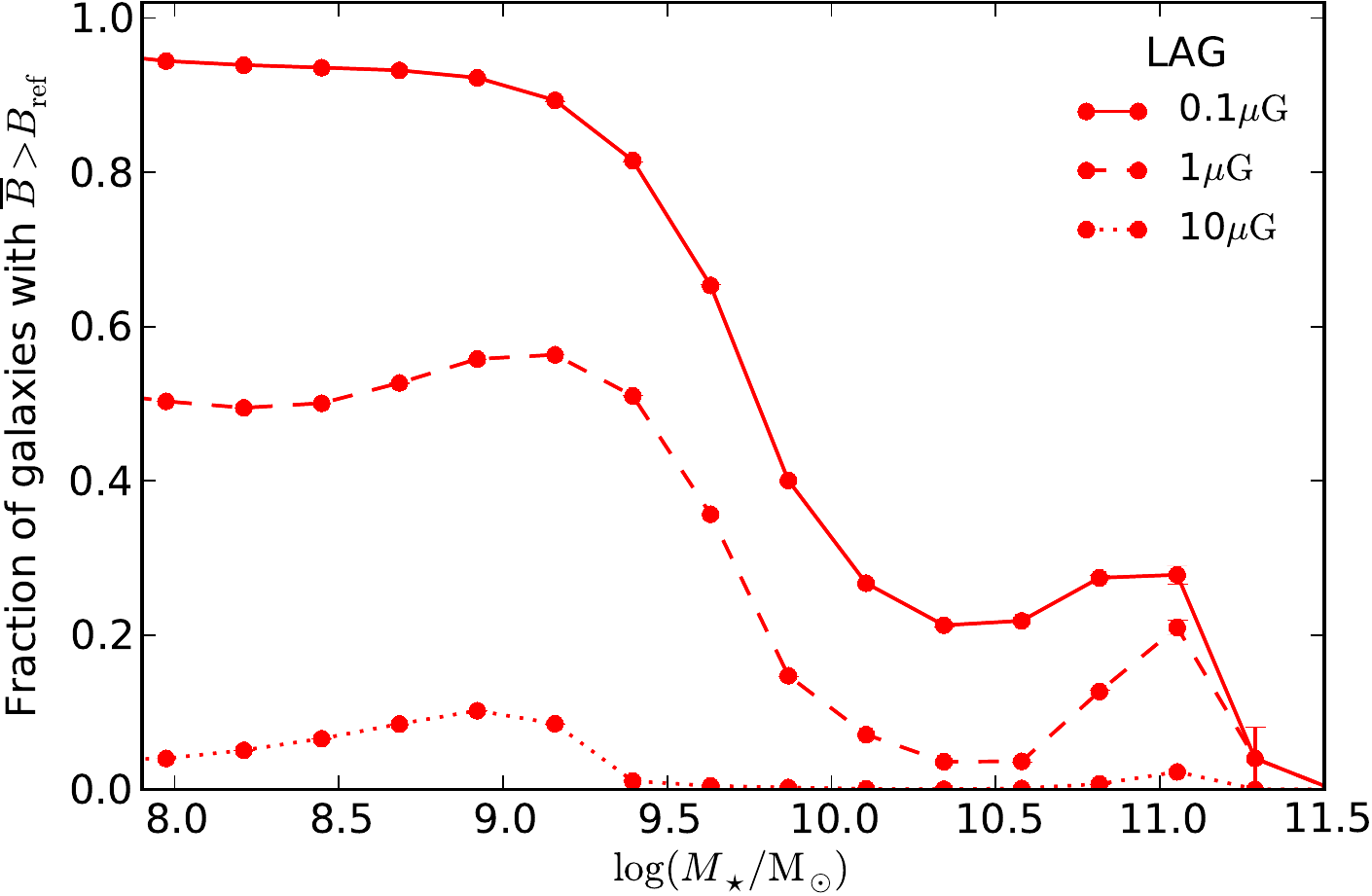}
\\
\includegraphics[width=\columnwidth]{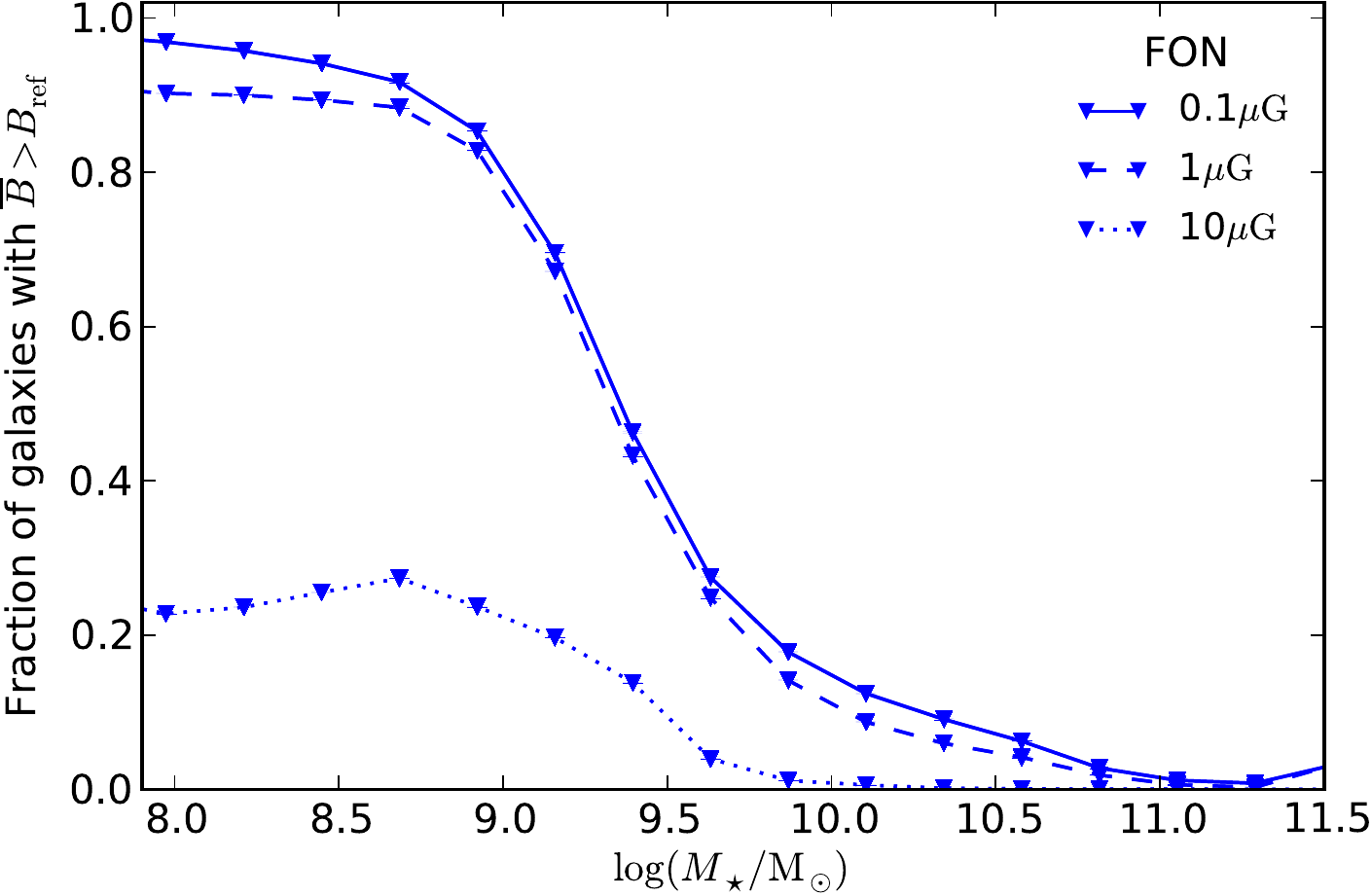}
 \caption{
          The fraction of galaxies with $\Breg>0.1$, $1$ and $10\muG$ at $z=0$
          as a function of the galaxy mass in each model, as specified in the
          legend.
         }
 \label{fig:fractions}
\end{figure}

In Fig.~\ref{fig:fractionsA}, we show the fraction of
galaxies with active large-scale dynamos (i.e., $|D|>|D\crit|$)
as a function of galaxy mass at $z=0$. The fraction
of active dynamos in galaxies \emph{decreases with mass\/} and all models
predict that fewer than 40 per cent of galaxies have active dynamos for
$M\gtrsim 10^{10}\msun$. Fig.~\ref{fig:fractions} provides a more
detailed picture where the fraction of galaxies with a magnetic
field strength exceeding $0.1,\ 1$ and $10\muG$ is shown as a function of mass
for each galaxy formation model separately.


\begin{figure*}
 \centering
 \includegraphics[width=\textwidth]{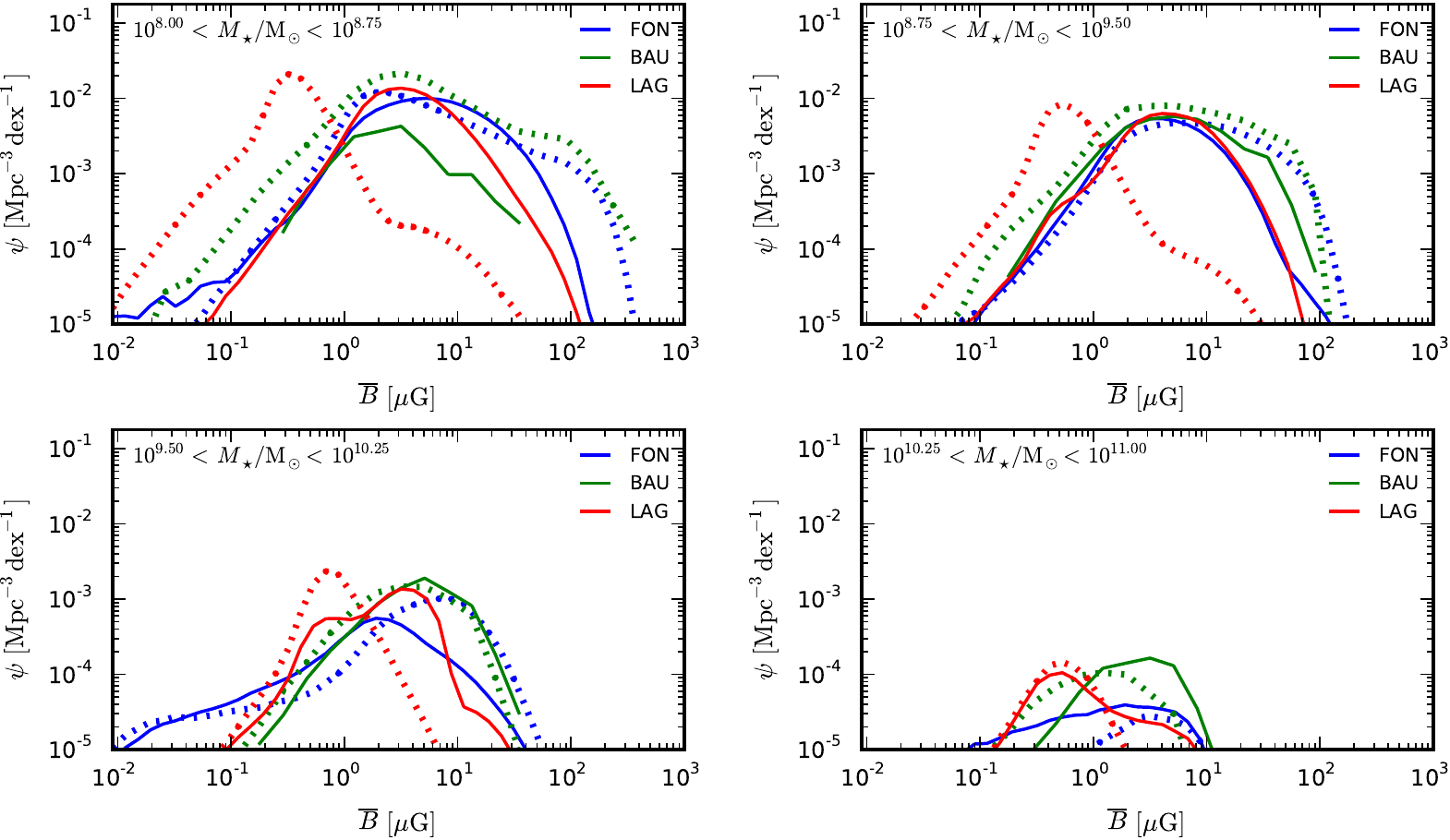}
 \caption{
 The distribution of large-scale magnetic field strengths at $z=0$; each panel
 shows different galaxy stellar mass interval, as labelled.
 Dotted curves show the MSF associated with satellite galaxies and continuous
lines show the MSF of central galaxies.
}
 \label{fig:msf_satcen}
\end{figure*}

The bimodal nature of the MSF of the LAG model in Fig.~\ref{fig:msf} is
clarified by Fig.~\ref{fig:msf_satcen} where the MSF is shown separately for
the central galaxies and their satellites: both the central galaxies and the
satellites have unimodal MSFs but with maxima at different values of $\Breg$
in the LAG model. This is a consequence of the differences in the gas content of
the
satellite and central galaxy populations {(see section~\ref{sec:gascontent} for 
details).}

\begin{figure*}
 \includegraphics[width=\textwidth]{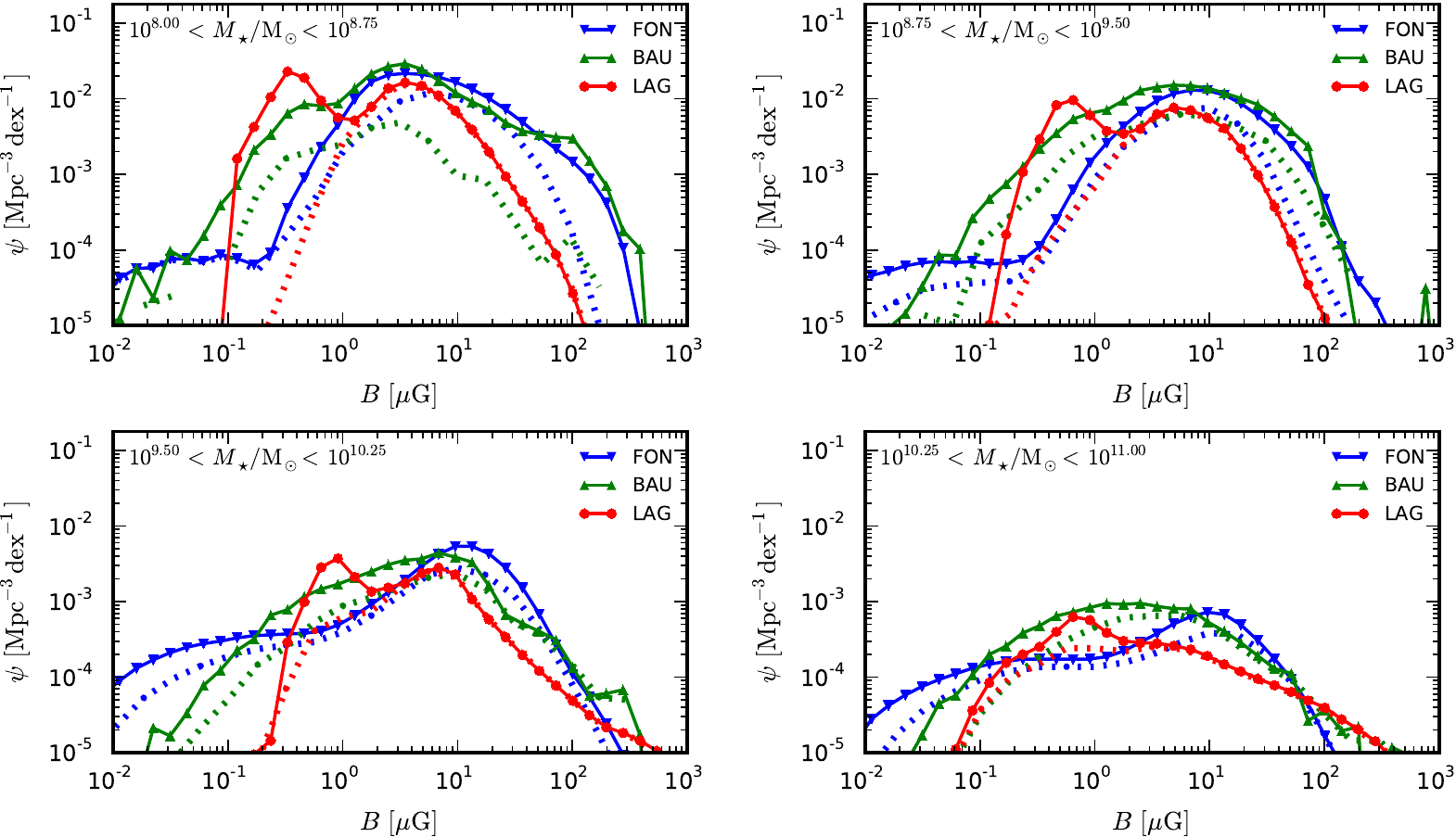}
 \centering
 \caption{
          The distribution of the total magnetic field strength,
          $B= \left(\overline{B}^2+b^2\right)^{1/2}$,
          in different galaxy stellar mass intervals at $z=0$, as labelled in
          each panel. Dotted lines show the corresponding distributions for the
          central galaxies alone. Different colours are used to indicate
          different SAMs as specified in the legend.
         }
 \label{fig:msf_total}
\end{figure*}

\subsection{The total magnetic field}
Fig.~\ref{fig:msf_total} shows the MSF of the total magnetic field comprising
both the large-scale and random parts,
$B = (\Breg^2+b^2)^{1/2}$; the MSF of the central galaxies
is also shown {with} a dotted curve.
The overall features of the large-scale magnetic field distributions of
Fig.~\ref{fig:msf} can be seen in the total field as well.
The most important differences occur at the two highest mass intervals where
the MSF of the total magnetic field is significantly broader than the MSF of
$\Breg$. At these masses, the small-scale magnetic fields dominate. This is
consistent with the small fractions of active large-scale dynamos
in these galaxies, as shown in Fig.~\ref{fig:fractions}.

In the LAG model, the tail of large-scale magnetic fields
$\Breg\lesssim0.1\mkG$ is concealed by stronger random magnetic fields,
producing a cut-off at a mass-dependent minimum field
strength.
On the other hand, in the FON model, the opposite happens: there is a
population galaxies (both satellites and central) with negligible large-scale
magnetic field which contains random magnetic fields $b\lesssim 0.2\muG$.
Such a change does not occur in the BAU model where the total magnetic
field has a distribution not dissimilar to that of the large-scale field alone.

\begin{figure*}
 \centering
 \includegraphics[width=\textwidth]{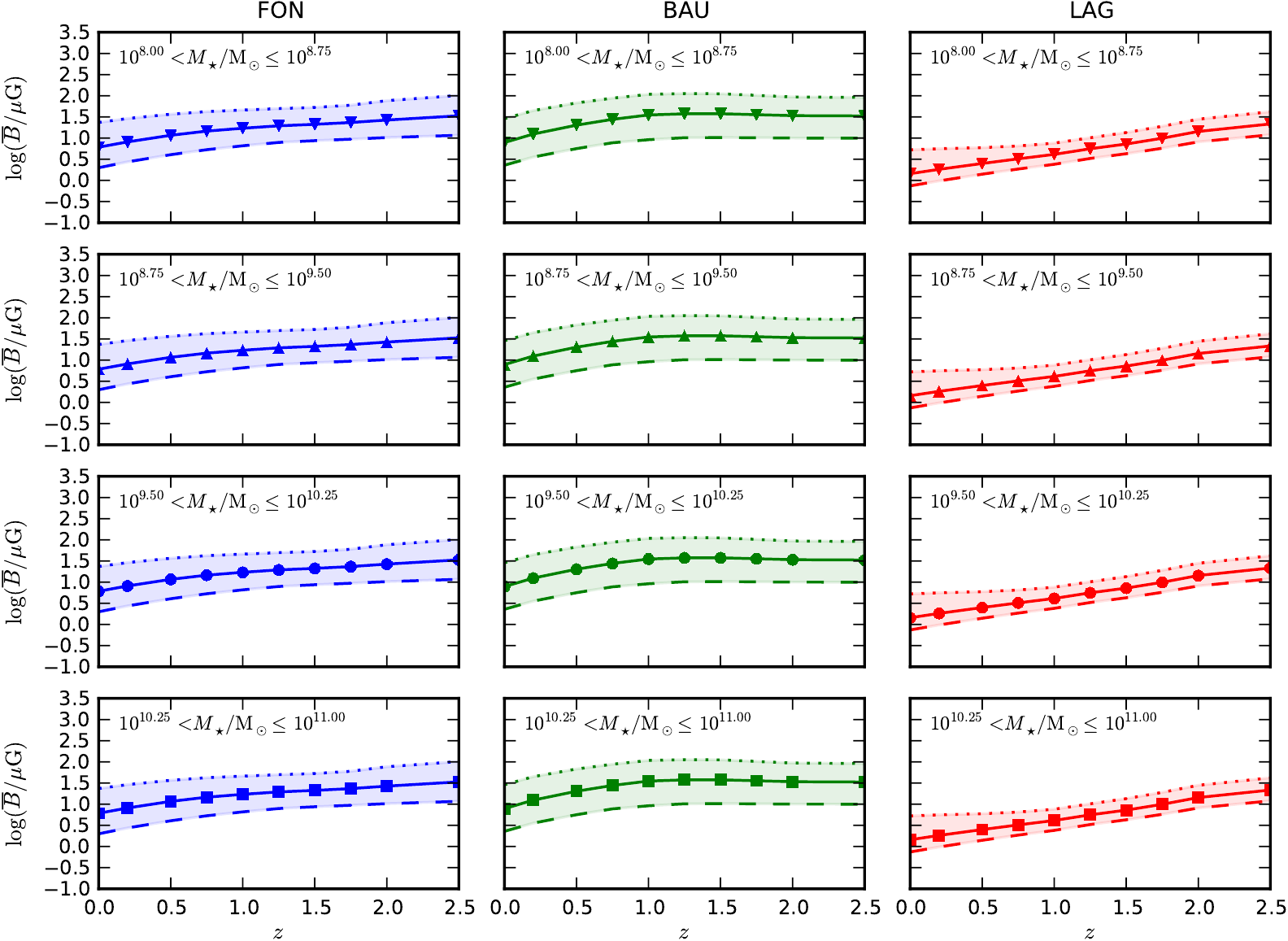}
 \caption{
          Redshift evolution of the large-scale magnetic field. The solid lines
          show the median field values. Dashed and dotted lines correspond to
          the 15 and 85 percentiles, respectively.
          Different columns correspond to different models (as indicated in the
          top row)
          and different rows show the results for different
          galaxy mass intervals (as indicated).
         }
 \label{fig:mode_evo}
\end{figure*}

\subsection{Evolution of galactic magnetic fields over cosmological time scales}
Fig.~\ref{fig:mode_evo} shows the redshift evolution of the median
value of the steady-state large-scale magnetic field in each galactic mass
interval, as well as the 15th and 85th percentiles.
Thus, at \emph{a fixed mass interval}, the median of the large-scale magnetic
field increases quasi-exponentially with redshift.
{The main reason for this increase are the larger interstellar gas densities
that
occur at higher redshifts.} 
One must, however, bear in mind that Fig.~\ref{fig:mode_evo} reflects
changes in both magnetic field strength and galaxy mass with redshift.

The large-scale magnetic fields are expected to evolve on a time scale of
$10^8\text{--}10^9\yr$ \citep{BPSS94,S07}. On the one hand, the dynamo time
scale
is shorter than the galactic evolution time, which allows us to use the
steady-state
strength of magnetic field in our estimates presuming that the dynamo action
generates the large-scale magnetic field in a young galaxy and then adjusts
itself
instantaneously to the evolving galactic environment. This approximation is
evidently acceptable in the inner parts of a galaxy where the rotational
velocity
shear is high and hence the dynamo time scale is short. However, the effects
of a finite dynamo time scale can be significant in the outer parts of galaxies.

{To illustrate these arguments, consider again the time scale of the 
mean-field dynamo given by equation~\eqref{tauMF}.
Assuming, for the sake of illustration and by analogy with the Milky Way, that 
$h=0.2\kpc$ at $r=1\kpc$ and $\Omega=V_0/r$ with $V_0=200\kms$, we obtain
for the inner galaxy
\[
\tau\MF\simeq20\Myr\left(\frac{r}{1\kpc}\right)\,,
\] 
having adopted $D\crit\approx-8$ and $|D|\simeq30\gg|D\crit|$ at $r\simeq1\kpc$.
Thus, the e-folding time of the large-scale magnetic field can be as short as
$20\Myr$ in the inner galaxy. With $h=0.5\kpc$ at $r=10\kpc$, and adopting 
$D\crit-D=4$ for illustration, we have  $\tau\MF\simeq0.7\Gyr$
{in the outer galaxy}.
The effective large-scale seed magnetic field produced by the fluctuation
dynamo
is estimated by  \citet{BPSS94} as $\Breg_\mathrm{s}=10^{-3}\mkG$; then the 
time required to amplify it to a strength $\Breg=1\mkG$ is about $7\tau\MF$.
These estimates suggest that the inner few kiloparsecs of galactic discs would 
host $\mu$G-strong large-scale magnetic fields very soon after their formation.
A galaxy formed at a redshift $z=10$ would have its central kiloparsec 
magnetized with a large-scale magnetic field by $z\approx8$, whereas 
a microgauss-strong magnetic  field can build up via the local dynamo 
action  by $z\approx1$ at $r=10\kpc$ \citep[see also][]{BPSS94}.}

{The above estimate is rather conservative since,
due to the stronger dynamo action in the inner galaxy, the outer radius 
of the magnetized region increases approximately linearly in time at a speed
$V_\mathrm{f}\simeq\sqrt{\gamma\eta}$ \citep{MSS98}; this expansion can be
described
as the propagation of a magnetic front driven by the mean-field dynamo action. 
Taking $h=0.3\kpc$ as a representative value for the whole disc 
and $\eta=10^{26}\cm^2\s^{-1}$, we obtain
$V_\mathrm{f}\simeq6(r/1\kpc)^{-1/2}\kms$,
and the front reaches $r=10\kpc$ in $t=\int_0^r\mathrm{d} r/V_\mathrm{f}\simeq
3.5\Gyr$.
Thus, a magnetic front propagating outwards from the inner galaxy can produce a
strong
magnetic field in the outer galaxy by $z\simeq1.5$, sooner than suggested by
the above estimate of the local growth time at $r=10\kpc$. Effectively, the
propagating
magnetic front provides a strong seed magnetic field for the local dynamo action
thus leading to a faster build-up of the large-scale magnetic field.}

{We stress that turbulent magnetic fields can be produced on much shorter time
scales anywhere in the galaxy (Section~\ref{sec:Bsmall}).}

{These estimates of the redshift at which a spiral galaxy can develop a
$\mu$G-strong
large-scale magnetic field are somewhat higher than those of 
\citet{Arshakian2009,Arshakian2011} since these authors did not consider the
fact
that the dynamo time scale is shorter in the inner galaxy: their estimates apply
to
the local dynamo action at $r\simeq10\kpc$.
}

{Since the dynamo model used here does not include the dynamo time scales
explicitly,
assuming that they are shorter than the galactic evolution times, we only extend
our
results to $z=2.5$.
At this redshift, the outer radius of the part of the galactic disc occupied by

the large-scale magnetic field of a microgauss strength is about $6\kpc$,
allowing
for local dynamo action alone. Extension of the model to include evolutionary
equations for magnetic field will be published elsewhere.}

\begin{figure*}
 \centering
 \includegraphics[width=\textwidth]{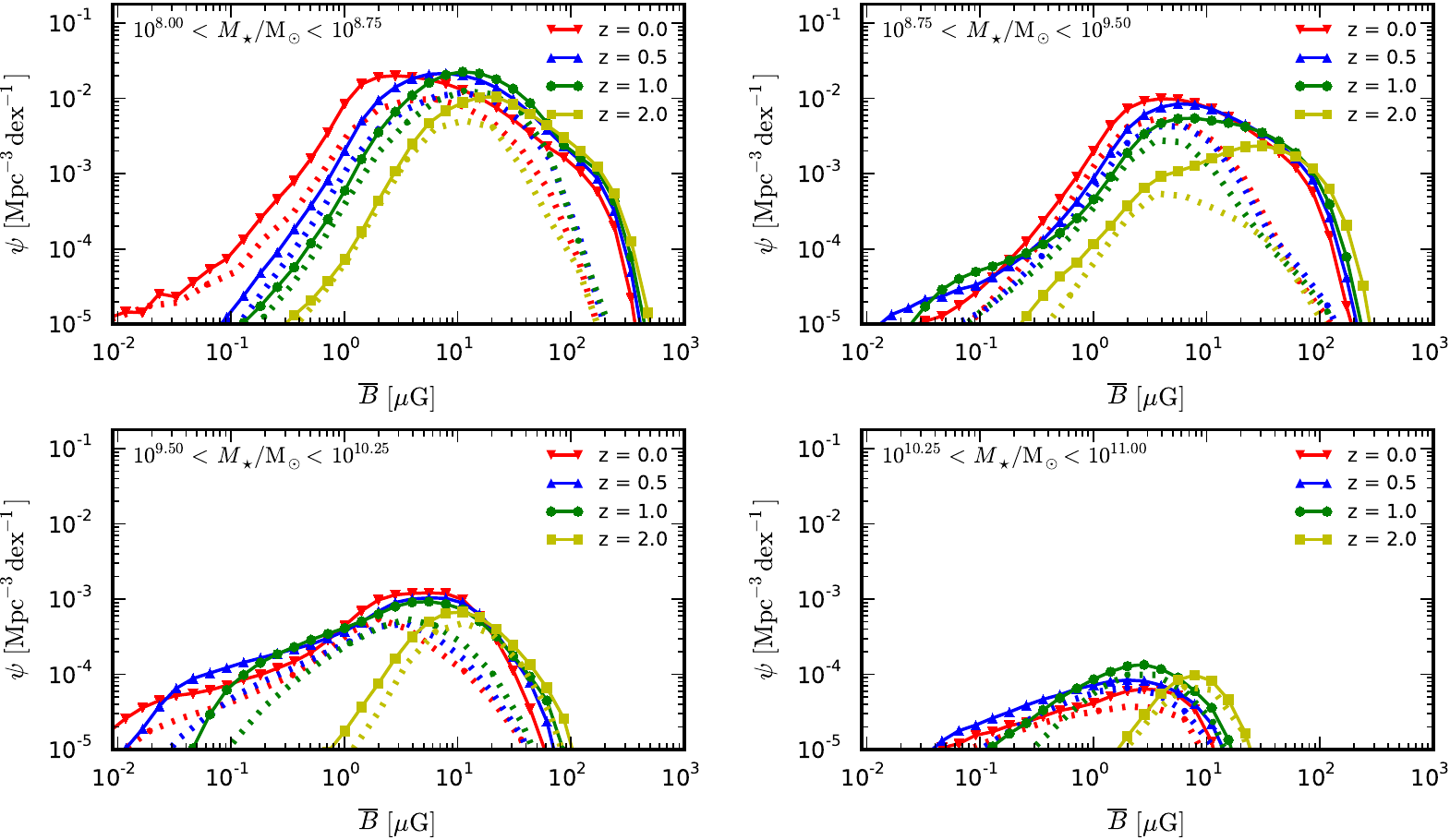}
 \caption{
          The evolution of the MSF in the case of the FON model.
          The dotted curves correspond the MSF of central galaxies.
          Different colours indicate the predictions for different redshifts as
          labelled.
 }
 \label{fig:msf_evo_fon}
\end{figure*}
\begin{figure*}
 \centering
 \includegraphics[width=\textwidth]{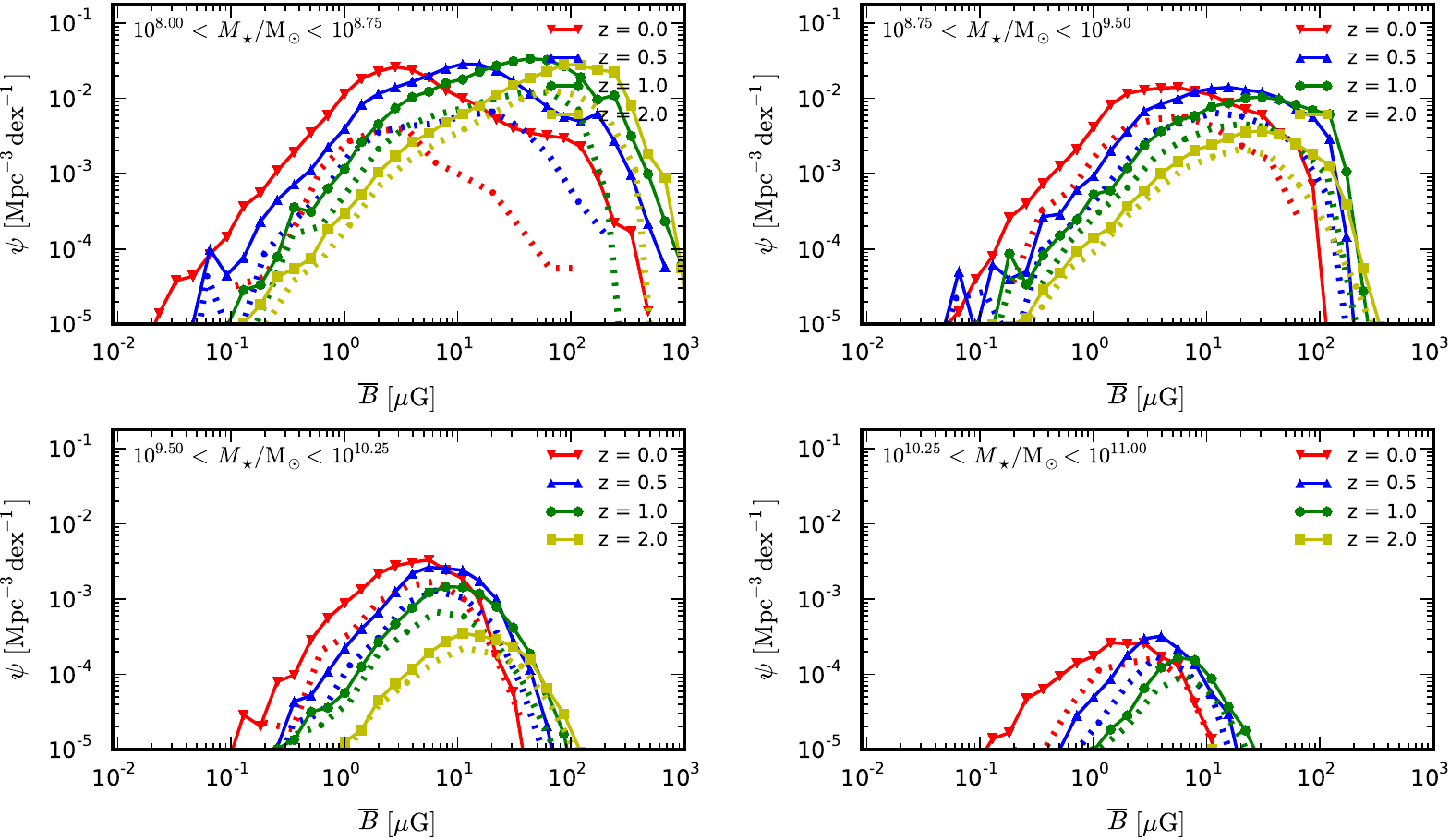}
 \caption{
          The evolution of the MSF for the BAU model.
          The dotted curves correspond the MSF of central galaxies.
          Different colours indicate the predictions for different redshifts as
          labelled.
         }
 \label{fig:msf_evo_bau}
\end{figure*}
\begin{figure*}
 \centering
 \includegraphics[width=\textwidth]{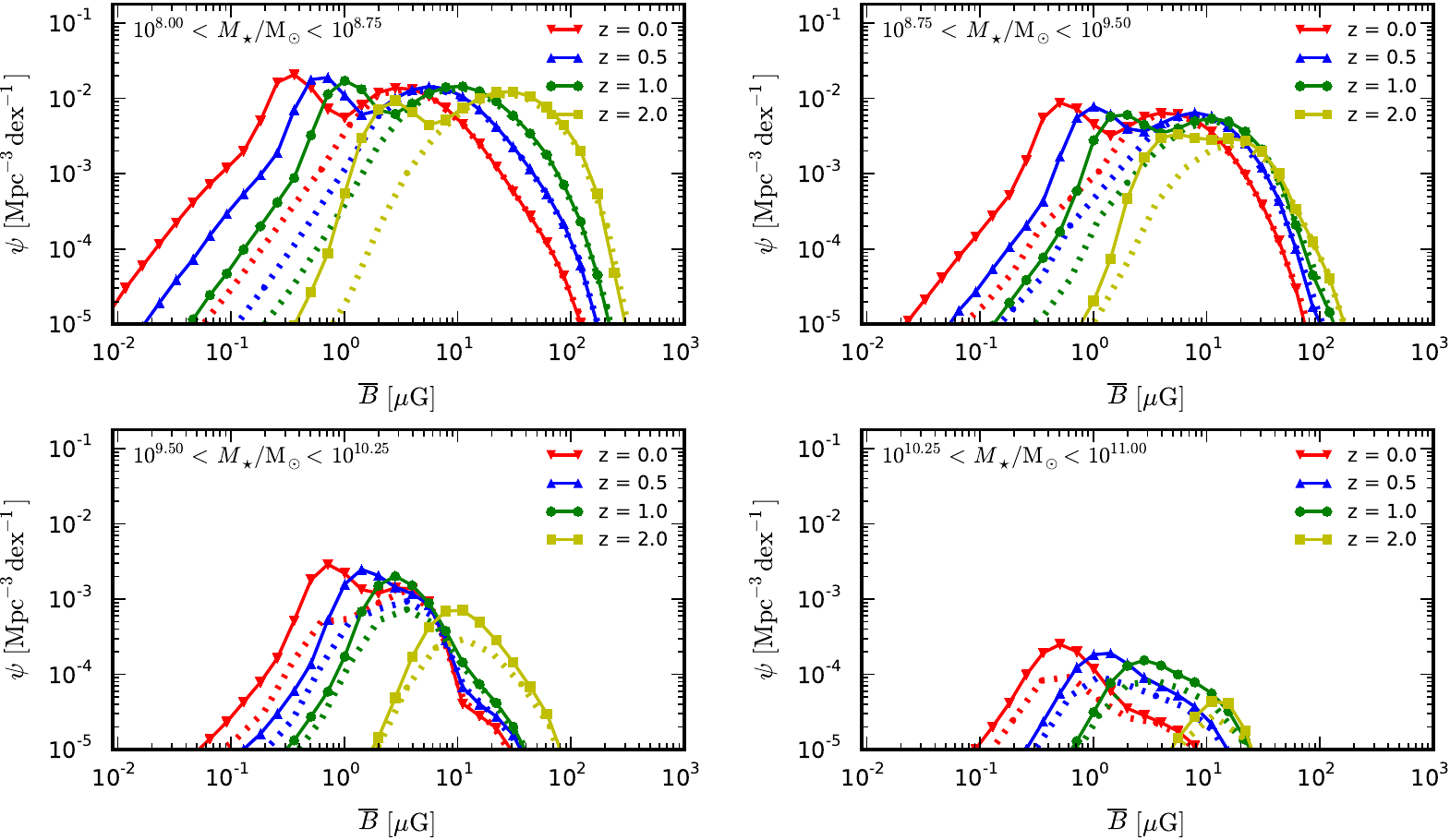}
 \caption{
          The evolution of the MSF for the LAG model.
          The dotted curves correspond the MSF of central galaxies.
          Different colours indicate the predictions for different redshifts as
          labelled.
}
 \label{fig:msf_evo_lag}
\end{figure*}


Figures \ref{fig:msf_evo_fon}, \ref{fig:msf_evo_bau} and
\ref{fig:msf_evo_lag} show the MSF at different
redshifts for each galaxy formation model. Except for a shift towards
higher values of $\Breg$, there is not much variation in the shape of the MSF
with redshift for the less massive galaxies.

In the same figures, dotted curves show the MSF of central galaxies alone,
suggesting that there is no significant change in the relative contributions of
the central and satellite galaxies to the MSFs.

\subsection{Magnetic fields and the gas content of a galaxy}
\label{sec:gascontent}
The form of the MSF is closely related to
the gas content of galaxies. The gas density enters the
reference magnetic field strength $B\eq$ through equation~\eqref{eq:Beq}.
Moreover, the efficiency of magnetic helicity advection, quantified
by $R_u$ of equation~\eqref{eq:Ru}, depends on the star formation rate,
which is controlled by the amount of cold gas in the galactic disc available
to be converted into stars.

Equation~\eqref{eq:B} shows that galaxies with stronger star formation,
and hence larger $R_u$, can have stronger large-scale magnetic fields. However,
larger values of $R_u$ also make the critical dynamo number
higher, as shown by equation~\eqref{eq:Dc}. As a result, there exists a
star formation rate optimal for the mean-field dynamo action. In galaxies
with large $R_u$ the dynamo action is suppressed. {Because of their higher star
formation rates, this affects massive galaxies especially strongly,
systematically
reducing, with the galactic mass, both the strength of their magnetic fields 
(as can be seen in the plots of the MSF) and}
the fraction of galaxies with active dynamos as shown in
Fig.~\ref{fig:fractionsA}.

\subsection{The importance of the ram-pressure stripping}

An important consequence of the non-linear relationship between the gas
content, star formation and the magnetic field strength is the difference
between the MSFs of central and satellite galaxies shown in
Fig.~\ref{fig:msf_satcen}. Because of the loss of hot gas by the satellites
due to ram-pressure stripping, both the gas density and star formation
rate in them decrease  leading to weaker outflows, thereby lowering $\Breg$.
The effect is more evident for the LAG model because of the assumption of
instantaneous ram-pressure stripping of the hot gas halo of the satellite (the
so-called satellite starvation).
For the FON model, the separation between the MSFs of satellite and
central galaxies is smaller, reflecting the gradual ram-pressure stripping
adopted in this model.

In the BAU model, despite the same efficiency of the ram-pressure
stripping as in the LAG model, the MSF has a maximum at nearly the same
positions
for both the satellite and central galaxy populations. The reason
is the star formation law adopted in this model, which leads to a
higher star formation rate in galaxies with smaller circular
velocity, as it follows from equation~\ref{eq:BAU_SFL}.
The increase in star formation rate in
lower-mass galaxies leads to an increase in $R_u$, compensating the effects of
starvation.

\subsection{Magnetic field strength and star formation rate}

The dependence of magnetic field strength on the SFR is due to an explicit
dependence of the large-scale field strength \eqref{eq:B} on the outflow
Reynolds number $R_u$ defined in equation~\eqref{eq:Ru}, with $v_\mathrm{ad}$
given in equations \eqref{eq:vad} and \eqref{eq:vout}. In addition, the SFR is
involved in the mean-field dynamo action through the dependence of the critical
dynamo number on $R_u$ in equation~\eqref{eq:Dc}. With the galactic outflow
model used here (Appendix~\ref{ap:OB}), the outflow speed is proportional to
$L_\mathrm{SN}^{1/3}$, the energy supply rate from supernovae in OB
associations, and thus to $\mathrm{SFR}^{1/3}$. However, we then assume that
all OB associations have the same energy supply rate, so that the outflow speed
is proportional to the SFR.
As a result, equation~\eqref{eq:B} predicts that
$\Breg\propto\mathrm{SFR}^{1/2}$
in galaxies with a strong outflow (where $R_u>\upi^2 R_\kappa$),
$\Breg$ is independent of the SFR if the outflow is weaker, and the dynamo
action is suppressed completely by very strong outflows, such that
$|D|<|D\crit|$.

\begin{figure*}
 \centering
\includegraphics[width=\textwidth]{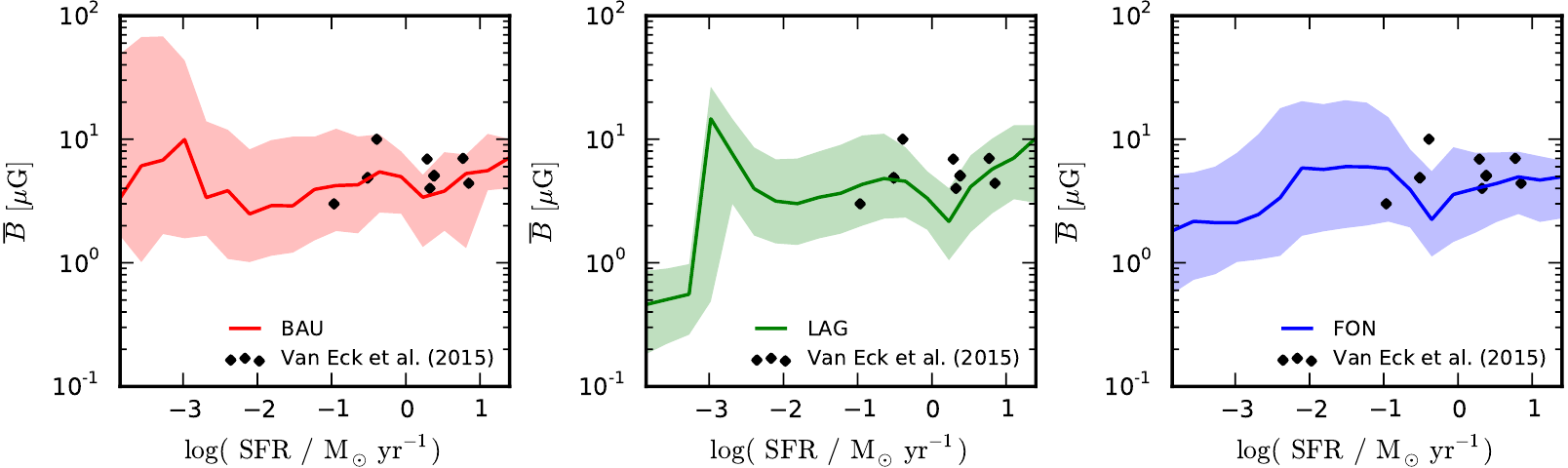}
\includegraphics[width=\textwidth]{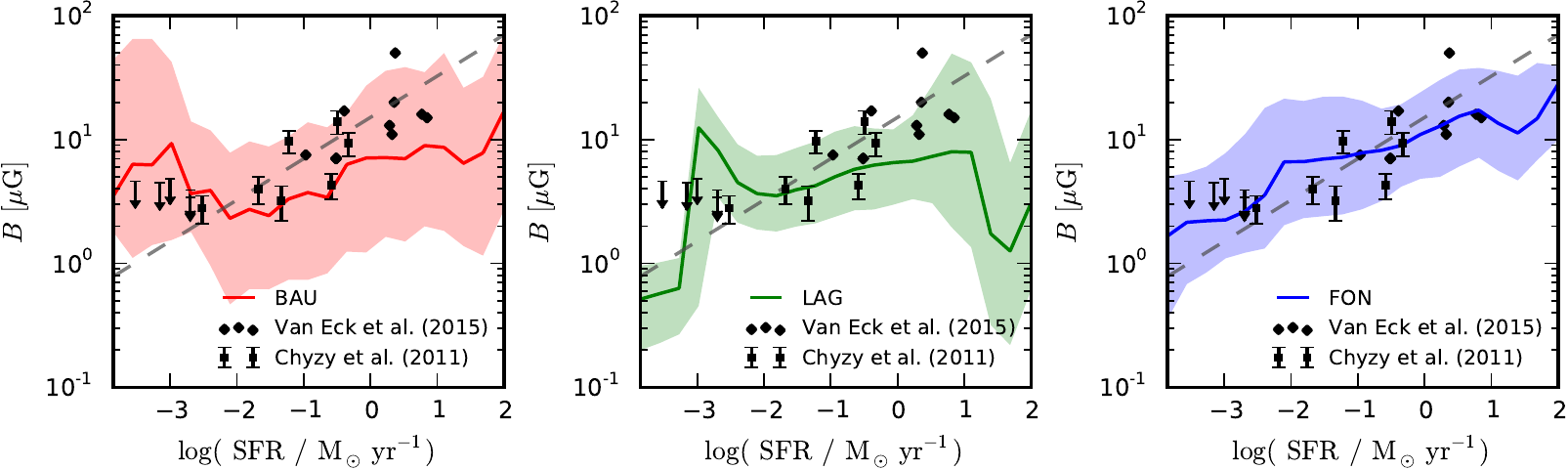}
 \caption{{The dependence of the large-scale field strength, $\overline B$
 (upper row), and total magnetic} {(lower row)}{, $B$, on the star
formation
 rate (SFR) predicted by each model. The solid lines show the median magnetic
field
 values while the shaded areas show the range between the 15th and 85th
 percentiles. The square data points }{and triangles correspond,
respectively,
 to the observed values and upper limits obtained} {
 by \citet{Chyzy2011} for dwarf irregular galaxies in the Local Group. The
 diamond-shaped data points were calculated from the compilation made by
 \citet[their Table 2]{vanEck2015}: when the surface density of
 star formation, $\Sigma_\text{SFR}$ was absent, we adopted the value
 corresponding to the innermost galactic radius; the total SFR was then obtained
 by integrating the surface density of SFR, $\Sigma_\text{SFR}$, and the
overall
 magnetic field was taken as the area-weighted average of the reported values. 
 The dashed grey curve in the bottom row shows the relation
 $B\propto\text{SFR}^{1/3}$
 with an arbitrary choice of the intercept (see text).}
}
 \label{fig:SFRvsB}
\end{figure*}

{
The overall dependence of the total and large-scale magnetic fields strengths
on the SFR obtained in our model is shown in the bottom and top
rows of Fig.~\ref{fig:SFRvsB}, respectively.
}
{
The large-scale magnetic field strength is roughly independent of the SFR and 
the model predictions are compatible with observations of nearby spiral
galaxies
compiled by \citet{vanEck2015}.
}
{
The second row shows the variation of total magnetic field strength with the
SFR.
The model predictions are compatible with observations of both  nearby spiral
galaxies and Local Group dwarf irregular galaxies \citep{Chyzy2011}.
}
{
For comparison, we also show (the dashed grey curve) the frequently adopted
power law
dependence $B\propto\text{SFR}^m$ with $m\approx1/3$. A power law dependence is 
compatible with the predictions of the BAU and LAG over narrow SFR ranges and of
the
FON model over the whole SFR range.
}

{
When interpreting the lower row of Fig.~\ref{fig:SFRvsB}, one ought to bear in
mind that, since the total magnetic field is dominated by the random component
and the turbulent velocity is fixed in our models (see Section
\ref{sec:Bsmall}),
the emerging relation between the field strength and the SFR is a consequence of
the non-linear relation between gas density and SFR and \emph{not} of the
dependence of the turbulent velocity on the SFR.
}

{\citet{SB13} argue that inverse Compton scattering of cosmic ray electrons
can dominate their synchrotron losses from redshift $z\approx3$ upwards,
thus affecting the observability of galactic magnetic fields at high redshift.
We note, however, that these authors use interstellar gas parameters
obtained by \citet{TQM05} for starburst galaxies with the surface density
of star formation $\Sigma_\star\simeq10^3\,\msun\kpc^{-2}\yr^{-1}$, where 
vertical support against gravity is provided by radiation pressure. 
However, they apply this model to galaxies with surface density of SFR
of order $\Sigma_\star=0.1\,\msun\kpc^{-2}\yr^{-1}$ where the properties of 
the interstellar gas are rather different.
Perhaps more importantly, \citet{SB13} employ a simple relation
$B\propto\Sigma_\star^{1/3}$ for the total magnetic field strength (dominated by
the
random magnetic field). As discussed above, any 
relation of $B$ to $\Sigma_\star$ is likely to be more complicated and can
definitely
be different in galaxies of different masses and different star formation
rates.
As discussed in Section~\ref{sec:IT}, the main reason for this is that the
turbulent
speed in the interstellar gas is self-regulated to be comparable to the sound
speed
(if the energy supply is high enough), because supersonic turbulence quickly
dissipates
the excess kinetic energy into heat thus increasing the speed of sound. In
galaxies
with strong star formation, a progressively larger part of the energy supplied
by
supernovae and stellar winds is channeled into galactic outflows rather than
turbulence.
Since the hot gas leaves the disc at time scales shorter than even the
fluctuation dynamo time, both large-scale and random magnetic fields of galactic
discs are generated in the warm gas where the speed of sound is of order
$10\kms$.
}
{
This strongly nonlinear effect seems to preclude any universal dependence of
the
turbulent speed and magnetic field strength on the SFR applicable beyond
relatively
narrow ranges of galactic mass and SFR. This makes the conclusions of
\citet{SB13}
questionable.
}
{The dependence of galactic magnetic field strength on star formation rate
is usually
discussed in connection with the radio--far-infrared correlation
\citep[e.g.][and references
therein]{YRC01}, where this dependence is just one element of a complex physical
system. It is
not our intention to discuss or explain that correlation, even though our
conclusions
are relevant to the discussions of its nature.}

{Since galaxy mergers are instantaneous in SAMs, we cannot include any
effects of the mergers on galactic magnetic fields. However, the starburst
produced
by a merger is a part of the SAMs; both large-scale and turbulent magnetic
fields are
enhanced during a starburst in our model. The effects of galaxy mergers on the 
mean-field dynamo can be nontrivial \citep{Drzazga2011,Geng2012,Kotarba2011}, 
but their inclusion would require a more detailed model of
both galaxy formation and dynamo action that those used here.}

\begin{table}
\caption{\label{tab:Ru} Fraction of galaxies with active dynamos and $R_u<\pi$
in the galaxy formation models explored.}
 \centering
 \newcommand{\mc}[3]{\multicolumn{#1}{#2}{#3}}
 \begin{tabular}{cccc}
\hline
Mass interval            &LAG &BAU &FON \\
$\log M_\star/\msun$    &percentage     &percentage     &percentage \\
\hline
    8.00--8.75   &85  &88  &93\\
    8.75--9.50   &89  &89  &95 \\
    9.50--10.25  &92  &87  &92 \\
    10.25--11.00 &92  &89  &90 \\
    \hline
 \end{tabular}
\end{table}

\subsection{Advection and diffusion of magnetic helicity}
Table~\ref{tab:Ru} shows that most galaxies with an active
mean-field dynamo have $R_u<\pi$. Thus, the strength of the large-scale
magnetic field, given in equation~\eqref{eq:B}, is dominated by
the diffusion of magnetic helicity for the value of $R_\kappa$
used, $R_\kappa=0.3$. This reduces significantly the effect
of galactic outflows on the mean-field dynamo action.

Equation~\eqref{eq:B} has been derived by \citet{Chamandy2014}
with the diffusive flux of the magnetic helicity
$\alpha_\mathrm{m}$ (proportional to the mean current helicity of the
random magnetic field, $\vect{b}\cdot\nabla\times\vect{b}$)
obtained assuming that the scale height of $\alpha_\mathrm{m}$ is equal to
that of the ionised gas. Almost nothing is known about the spatial
distribution of the mean current helicity even from  numerical simulations,
not to mention observations. The numerical simulations of \citet{GZER08}
\citep[see also][]{GBE13} suggest that the scale of the total $\alpha$-effect
across the gas layer can be significantly larger (at about $1\kpc$ in the
Solar vicinity of the Milky Way) than the gas scale height,
perhaps due to the contribution of $\alpha_\mathrm{m}$. If this is the case, the
value of $R_\kappa$ is strongly overestimated. Then the dynamo model
underestimates
the role of galactic outflows and, hence, star formation on the large-scale
magnetic
field. The relative roles of galactic outflows and diffusion in the
nonlinear mean-field dynamos requires further careful study under realistic
conditions for spiral galaxies.


\section{Summary and conclusions}
\label{sec:conclusions}

By coupling a nonlinear mean-field dynamo model with three well-established
semi-analytic models of galaxy formation, we have developed a framework to
predict the
strength of large-scale (global) and small-scale (turbulent) magnetic fields in
evolving spiral galaxies from the time when their bulge to disc mass ratio
reduces below 1/2 and they are classified as late types. We present our results
for various galactic stellar mass ranges selected
according to the physical properties of the galaxies and we consider the
evolution of  galactic magnetic fields with redshift. Our main assumption is
that the steady-state strength of the magnetic field is established
instantaneously as the host galaxy evolves. This assumption is not questionable
for the turbulent magnetic fields as the time scale of the fluctuation dynamo
producing them can be as short as $10\Myr$. The time scale of the large-scale
(mean-field) dynamo is longer at about a few Gyr
in the main part of a
spiral galaxy but can be {two orders} of magnitude shorter in the inner
galaxy. This time scale is {at least} marginally shorter than the galactic 
evolution time, which justifies our assumption.

We find that the choice of the galaxy formation model strongly affects
the number density of galaxies that host a large-scale magnetic
field of a given strength (called the MSF in the text) in most mass ranges.
In other words, the probability distribution of the strength of the large-scale
magnetic field {in a representative sample of galaxies} is sensitive to the 
galaxy formation model. We discuss how
the shape of the MSF is related to the physical processes affecting the dynamo
action and how these depend on the galaxy formation model.
In particular, the ram-pressure stripping of the gas in the haloes of satellite
galaxies results in quite different typical field strengths in satellites and
their central galaxies, which can lead to a pronounced bi-modality of the MSF.
Our experience thus demonstrates the possibility of using observations
of galactic magnetic fields to constrain galaxy formation models.

Our results are also relevant to planning observational studies of galactic
magnetism
at high redshifts. Galaxies that host a large-scale magnetic field can produce
a
signal in polarised synchrotron emission, even when they are unresolved: the
degree
of polarisation depends on the orientation of the galactic disc to the line of
sight and the degree of order in the magnetic field \citep{Stil:2009}. Thus,
radio continuum surveys that cover both total and polarised synchrotron
emission, using the SKA and other radio telescopes, will be able to produce
statistical distributions of fractional polarisation for the local universe and
for different redshifts, that can be directly compared to our (or similar)
models.
This data
will provide a new window through which to test both galaxy evolution models and
theories of magnetic field evolution in galaxies.
For example, all of the galaxy formation models considered here predict that
less
than $50$ per cent of the spiral galaxies with stellar mass greater than 
$10^{10}\msun$ should host large-scale magnetic fields greater than $1\muG$ in 
strength. Similar statements are made for other mass ranges in the main text
and
illustrated in Fig.~\ref{fig:fractions}.

Our results suggest that the strength of the large-scale galactic magnetic field
increases nearly exponentially with redshift in all mass ranges. This does not,
however, mean that the magnetic field in an individual galaxy evolves
exponentially because individual galaxies migrate between the mass ranges and,
more importantly, individual galaxies become dynamo active at different stages
of their evolution (even within a relatively narrow mass interval) because of
different individual merger trees.

Our approach can be improved in several ways. Most importantly, by allowing for
a finite magnitude of the dynamo time scale, we shall be able to include
detailed evolution of magnetic field allowing us to predict with confidence the
appearance of individual galaxies in the radio range and to assess the
importance of magnetic fields effects on star formation in evolving galaxies,
and to trace back a given magnetic configuration to earlier stages.

\section*{Acknowledgements}
LFSR thanks Durham University for the hospitality during his visit and
acknowledges support from the European Commission's Framework Programme 7,
through the Marie Curie International Research Staff Exchange Scheme LACEGAL
(PIRSES-GA-2010-269264) and support from the Brazilian Agency CNPq
(202466/2011-6) during the initial part of this work.
LFSR, AS and AF have been supported by the Leverhulme Trust
(grant RPG-097) and STFC (grant ST/L005549/1).
CMB acknowledges support from the STFC grant ST/L00075X/1.
This work used the DiRAC Data Centric system at Durham University, operated
by the Institute for Computational Cosmology on behalf of the STFC DiRAC HPC
Facility (www.dirac.ac.uk). This equipment was funded by BIS National
E-infrastructure capital grant ST/K00042X/1, STFC capital grant ST/H008519/1,
and STFC DiRAC Operations grant ST/K003267/1 and Durham University. DiRAC is a
part of the National E-Infrastructure.
{We are grateful to the referee for useful comments and suggestions.}

{ \footnotesize
\bibliography{sam_avg_magnetic_fields}
}
\bsp
\appendix

\section{The scale height of a pressure-supported thin disc}
\label{ap:h}
\glf assumes that the stars and gas in galaxy discs have
an exponential radial profile of the stellar surface density,
\begin{equation}
 \Sigma_{\star/\mathrm{g}}(r) = \frac{M_{\star/\mathrm{g}}}{2\pi r_\mathrm{s}^2}
\ee^{-r/r_\mathrm{s}}\,,\label{eq:Sigma}
\end{equation}
where $r_\mathrm{s}$ is the scale length and this is assumed to be the same for
stars
and gas.

Under the assumption of hydrostatic equilibrium in the $z$-direction, one
obtains (using cylindrical coordinates) the following expression for the gas
pressure,
$P$,
\begin{equation}
 \frac{\partial\,}{\partial z}\left(\frac{1}{\rho_\mathrm{g}}\frac{\partial
P}{\partial
z}\right) = 4\pi G\rho_\mathrm{t}\,,\label{eq:hydrostatic}
\end{equation}
{where $\rho_\mathrm{t}$ is the total gravitating mass density.}
Defining
{the surface densities}
\begin{equation}
 \Sigma(R,z) {=} 2 \int^z_0 \rho(R,z)\,\mathrm{d}z\,,
 \qquad
 \frac{\partial\Sigma}{\partial z} = 2 \rho(R,z)\,,\label{eq:derivSigma}
\end{equation}
which is related to the total surface density through
$\Sigma(R)~\!=~\!\lim_{z \to\infty} \Sigma(R,z)$ and using
equation~\eqref{eq:hydrostatic}, we obtain
\begin{equation}
  \frac{\partial P}{\partial z} = \pi G \Sigma_\mathrm{t} 
  	\frac{\partial\Sigma_\mathrm{g}}{\partial z}\label{eq:partial}\,.
\end{equation}
Since stars and gas are assumed to be distributed similarly, we have
$\Sigma_\mathrm{g} = k\Sigma_\mathrm{t}$ with $k$ a constant at a given time, 
which leads to
\[
\Sigma_\mathrm{t} \frac{\partial \Sigma_\mathrm{g}}{\partial z}
= \tfrac{1}{2} k\frac{\partial \Sigma_\mathrm{g}^2}{\partial z}\,.
\]
Then, integrating equation~\eqref{eq:partial}, gas pressure follows as
\begin{equation}
 P = \frac{\pi}{2} G \Sigma_\mathrm{g} (\Sigma_\mathrm{g} 
 		+ \Sigma_\star)\label{eq:Pnovo}\,.
\end{equation}
On the other hand,
\begin{equation}
 P = \frac{\alpha}{3} v_0^2 \rho \label{eq:P}\,,
\end{equation}
where $\alpha\ (>1)$ allows for various contributions to the interstellar
gas pressure: thermal, turbulent 
{and from} cosmic rays and magnetic fields.

The scale height is obtained by combining equations~\eqref{eq:Sigma},
\eqref{eq:Pnovo}, \eqref{eq:P} and $\rho \approx \Sigma/h$, i.e., assuming
weak vertical density variations:
\begin{equation}
 h(r)=\frac{\alpha v_0^2 r_\mathrm{s}^2}{G(M_\star + M_\text{gas})} \,
 	\ee^{r/r_\mathrm{s}}\,.
\end{equation}

\section{Galactic outflows driven by supernovae}
\label{ap:OB}
We follow equation (2) from \citet{MacLowMcCray1988} for the radius,
$R_\mathrm{sb}$,
of 
{a superbubble} produced by an OB association,
\[
 R_\mathrm{sb} = {267\,\text{pc}} \left(\frac{L_{38} \,
t_7^{\,3}}{n_0}\right)^{1/5}\label{eq:R}\,,
\]
{where $L_{38}$ is the mechanical luminosity in the unit of
$10^{38}\erg\s^{-1}$, $t_7$ is time in the unit of $10^7\yr$ and
$n_0$ is the number density of the ambient gas.}
The 
{superbubble expands at the speed}
\[
\dot R_\mathrm{sb} = 
{15.7\kms} L_{38}^{1/5} n_0^{-1/5} t_7
^{-2/5}\label{eq:dotR}\,.
\]
{\citet{MacLowMcCray1988} find} that the superbubble
breaks out
{from the disc}
when $R\approx 2h$. This can be used with the previous equations to 
{obtain the supperbubble age at the breakout},
\[
 t_{7}^{-2/5} = \left(\frac{n_0}{L_38} \right)^{-2/15} h_{134}^{-2/3}\,,
\]
{where $h_{134}=h/134\pc$.}
Therefore, the velocity of the shock front 
{produced by the superbubble at the break out} is
\begin{align*}\label{eq:vadsimple}
v_\text{sh} &= \left.\dot R_\mathrm{sb} \right|_{R_\mathrm{sb}=2h}  \nonumber\\
	&= {4\,\kms}
			\left(\frac{L_\mathrm{SN}}{10^{38} \,\text{erg s}^{-1}}\right)^{1/3}
			\left(\frac{n_0}{1 \,\text{cm}^{-3}}\right)^{-1/3} 
\left(\frac{h}{1\,\text{kpc}}\right)^{-2/3},
\end{align*}
and assuming $L_{38}=1$, one arrives at equation \eqref{eq:vadsingle}.

\label{lastpage}
\end{document}